\documentclass[12pt]{article}
\def\half{{\scriptstyle{\frac{1}{2}}}}

\usepackage{graphicx}
\begin{document}
\begin{center}
{\Large\bf Some thoughts on dynamic effective properties\\
-- a working document}\\
\vskip .25in
{\sl J.R. Willis, DAMTP, Cambridge}
\end{center}
\vskip .5in
The main purpose of this work is to address the question of the utility of ``effective constitutive
relations'' for problems in dynamics. This is done in the context of longitudinal shear waves
in an elastic medium that is periodically laminated, with attention restricted to plane waves
propagating in the direction normal to the interfaces. The properties of such waves can be found
by employing Floquet theory, implemented via a ``transfer matrix'' formulation. Problems occur
at frequencies beyond those that define the first pass band, associated in part with the difficulty
of assigning a unique wavenumber to the wave. This problem is examined, paying careful attention
to the requirements of causality and passivity. The transmission of waves into a half-space is discussed by studying the impedance of the half-space, both directly and in the ``effective medium''
approximation, and an alternative way of looking at this problem, based on construction of the Green's function, is developed. 
\section{Introduction}
The construction of effective relations for the dynamics of composites displays some difficulties but these have largely been resolved. The general form of the effective
relations is by now well-established. There is a complication associated with lack
of uniqueness but this is understood and under control. In particular, it is known
that one definition of the effective properties, which defines them uniquely, is to
insist that they should apply in the presence of any prescribed ``inelastic''
deformation, such as induced by temperature change or plastic deformation. Furthermore,
it is these properties that are delivered naturally and automatically from a formulation
involving a comparison medium.

There remains, however, an important question relating to the utility of an effective
medium formulation. It is (in principle) exact if it is expressed in terms of a Green's
function for the exact body, and type of boundary conditions, to which is is to be
applied, but finding the Green's function is tantamount to solving the original problem
which is not helpful. The usual procedure is to formulate the effective medium problem
for an infinite body in the hope that replacement of a finite composite medium by
one with the ``infinite medium'' effective properties will provide a useful approximation.
This is a valid approach in the ``homogenization limit'' in which the scale of the
microstructure is much smaller than any wavelength contained in the mean field that
is generated: a rigorous proof in the context of electromagnetic waves has been
given by Kohn and Shipman \cite{KS07}. But what if this restriction is not met? This question
has recently been addressed, through explicit study of a particular case, by Srivastava
and Nemat-Nasser \cite{SNN13}. The present note offers further discussion along similar lines.

The problem to be addressed is for a half-space occupying $x>0$, whose elastic constants and
density are piecewise-constant periodic functions of $x$. Srivastava and Nemat-Nasser \cite{SNN13}
adjoined a medium occupying $x<0$, which was taken to be ``effective medium''. They noted that,
if the half-space $x>0$ could reasonably be modelled as effective medium, there would be
little reflection of a wave incident on the interface from the medium occupying $x<0$.
In the sections that follow,
the problem is considered more generally, by comparing the impedance of the actual
half-space with its ``effective medium'' approximation.

\section{Waves in a periodic medium}

Consider an $n$-phase infinite periodic medium. Phase r occupies $x_{r-1}<x<x_r$ and has elastic
constant $E_r$ and density $\rho_r$, for $1\leq r\leq n$. The medium is repeated periodically
with period $h = x_n -x_0$. Define also $h_r = x_r-x_{r-1} = p_r h$, so that $p_r$ is the volume
fraction of phase $r$.

The equation of motion, Laplace transformed with respect to time, is
\begin{equation}
d\sigma/dx = sp,
\end{equation}
where stress $\sigma$ and momentum density $p$ satisfy the constitutive relations
\begin{equation}
\sigma = E\,du/dx,\;\;\;p = \rho v \equiv \rho\,(su)
\end{equation}
in the transform domain, where $u$ represents displacement. Thus, $u$ satisfies
\begin{equation}
(d/dx)\{E(x)du/dx\}-s^2\rho(x)u(x) = 0.
\end{equation}
The coefficients in this equation are periodic functions with period $h$. Hence it admits
two linearly independent solutions of the form
\begin{equation}
u(x) = e^{\mu x}\phi(x),
\end{equation}
where $\phi$ is periodic with period $h$.

In the case of the $n$-phase medium, $u(x)$ has the form
\begin{equation}
u(x) = A_r\cosh[k_r(x-x_{r-1})]+ B_2\sinh[k_r(x-x_{r-1})]\;\hbox{ for }\;x_r < x < x_{r+1},
\end{equation}
where
\begin{equation}
k_r = s/c_r,\;\;\;c_r = \sqrt{E_r/\rho_r}.
\end{equation}
The displacement $u$ and stress $E\,du/dx$ are continuous at all interfaces,
and 
\begin{equation}
u(x_n) = e^{\mu h}u(x_0)\,\hbox{ and }\;\sigma(x_n)=e^{\mu h}\sigma(x_0).
\end{equation}

The continuity conditions at $x_r$ for $1\leq r\leq n-1$ are satisfied by taking
\begin{equation}
\left(\matrix{A_{r+1}\cr
              B_{r+1}\cr}\right) = M_r\left(\matrix{A_r\cr
                                                    B_r\cr}\right),
\end{equation}
where
\begin{equation}
M_r = \left(\matrix{\cosh(k_rh_r)&\sinh(k_rh_r)\cr
                    (Z_r/Z_{r+1})\sinh(k_rh_r) & (Z_r/Z_{r+1})\cosh(k_rh_r)\cr}\right)
\end{equation}
with the impedance $Z_r$ of phase $r$ given by
\begin{equation}
Z_r = E_rk_r/s = \sqrt{E_r\rho_r}.  
\end{equation}
Conditions (2.7) then require that
\begin{equation}
(M-e^{\mu h}I)\left(\matrix{A_1\cr
                           B_1\cr}\right) = 0,
\end{equation}
where
\begin{equation}
M = M_1M_2\cdots M_n.
\end{equation}

The determinant of $M_r$ is $Z_r/Z_{r+1}$, with $Z_{n+1}=Z_1$. Hence, the determinant of
$M$ is $1$, and the equation for the eigenvalues $e^{\mu h}$ reduces to
\begin{equation}
\cosh(\mu h) = \half{\rm trace}(M).
\end{equation}
In the case $n=2$ this formula reproduces the dispersion relation of Rytov (1956).

Since inversion of the Laplace transform involves the time factor $e^{st}$, the wave travelling
in the positive $x$-direction is characterized by the eigenvalue for which ${\rm Re}(\mu) < 0$
when $s$ is real. More generally, it is a requirement of causality that ${\rm Re}(\mu) < 0$
when ${\rm Re}(s) >0$. This restriction on $\mu$ ensures that the wave amplitude remains
bounded as $x\to +\infty$.

While there is no ambiguity in the definition of the eigenvalue $e^{\mu h}$, $\mu$ itself
is not uniquely determined because if $\mu$ is a solution then so is $\mu + 2m\pi i/h$, for
any integer $m$. This has no effect on the actual solution but it does raise a problem
in relation to an effective ``dispersion relation''. Figure 1 shows a plot of values of $\mu h$
against frequency $f$, when $s=-(2\pi f)i+\epsilon$ (with $\epsilon$ small and positive),
as calculated directly using Matlab for an example two-phase medium which is discussed in more
detail later. The imaginary part of $\mu$ is denoted $q$.
\begin{figure}
\centering
\includegraphics[scale=0.5]{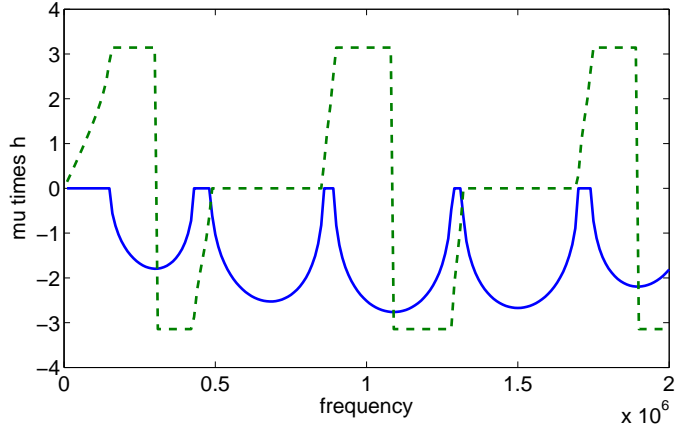}
\caption{\small Matlab-generated $\mu h$. Solid line: real part, dashed line: imaginary part.}
\end{figure}
\begin{figure}
\centering
\includegraphics[scale=0.35]{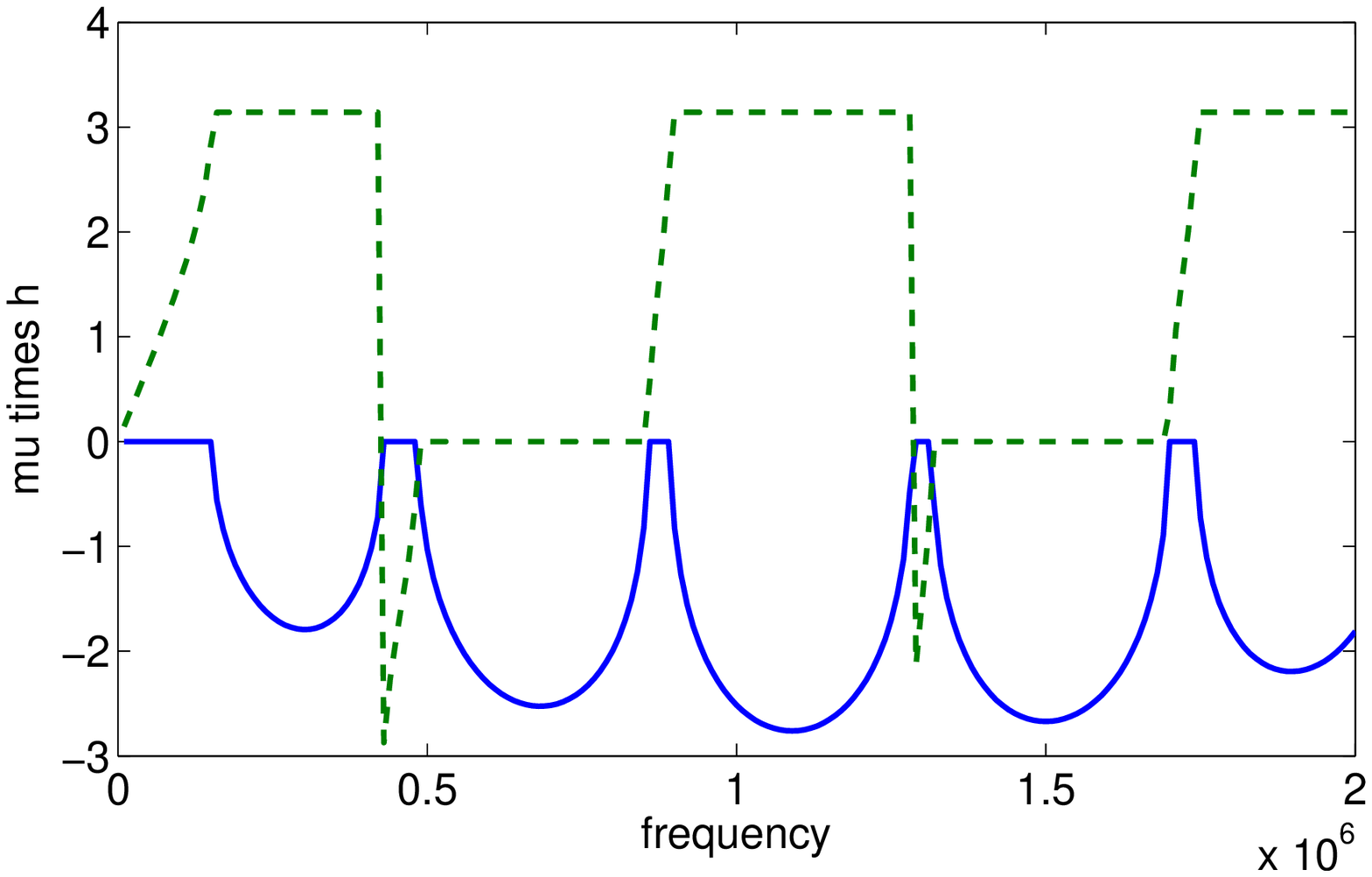}
\includegraphics[scale=0.35]{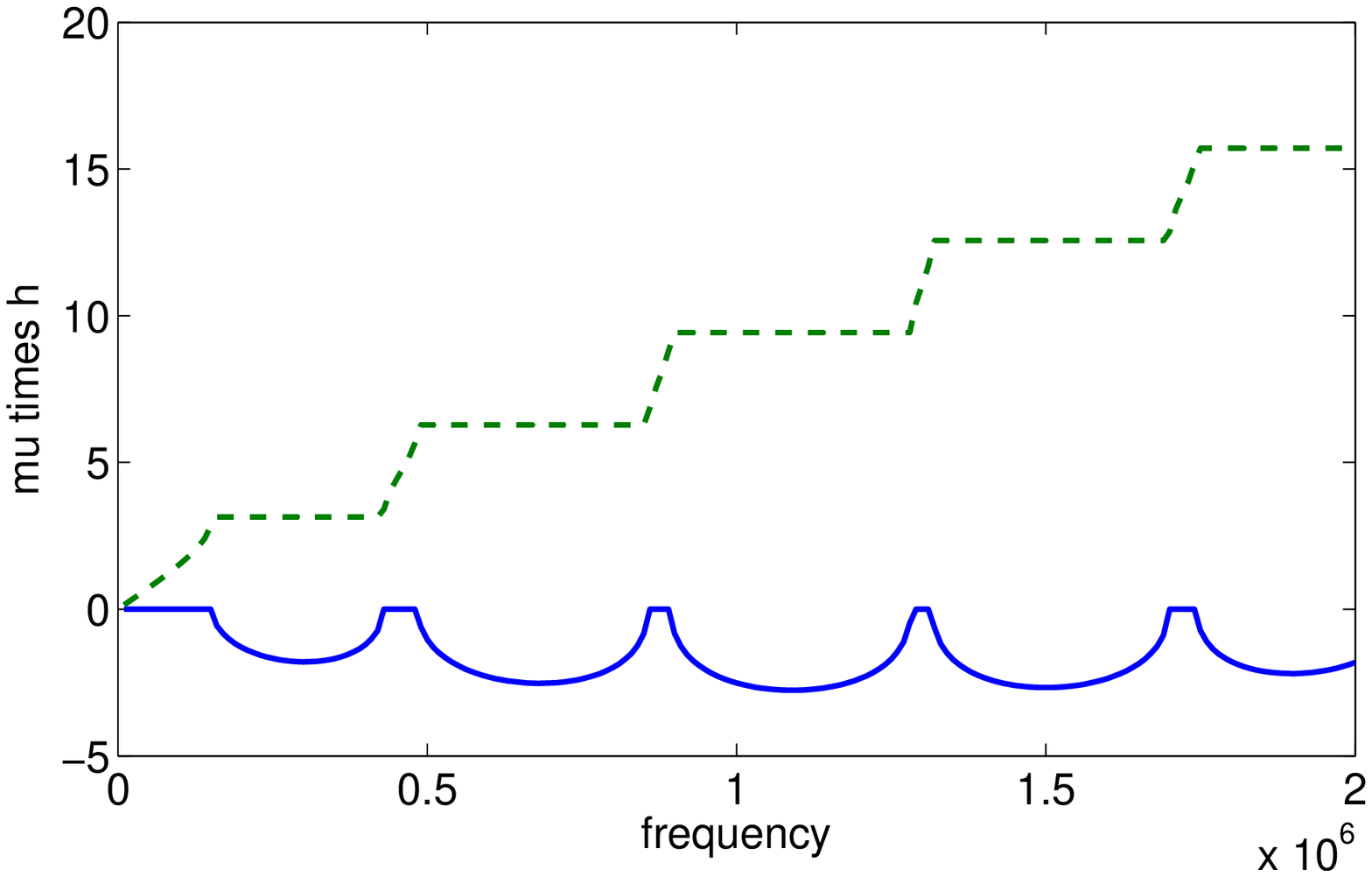}\\
(a)\hskip 6cm (b)
\caption{\small Two alternative forms for $\mu h$. (Note the different vertical scale in case (b).)}
\end{figure}
Figure 2 shows two alternatives. In the first, Fig. 2a, $qh$ is restricted
to the interval $(-\pi,\pi]$. In the second, Fig. 2b, $qh$ is continuous. Since the corresponding
time-dependence is $e^{-i(2\pi f) t}$, the group velocity is positive in the pass bands and the same
for all forms of $\mu$. Phase velocity depends on the choice of $\mu$ however. Perhaps the ``correct''
choice will depend on the psychology of perception and the importance assigned to ``aliasing''.

\section{Effective properties}
\setcounter{equation}{0}
The general form of the dynamic effective relations of a composite is
\begin{equation}
\sigma^{eff} = E^{eff}e^{eff} +  S_1 v^{eff},\;\;\;p^{eff} = S_2e^{eff} + s\rho^{eff}v^{eff},
\end{equation} 
where $e^{eff} = du^{eff}/dx$ and $v^{eff} = su^{eff}$. $E^{eff}$, $S_1$, $S_2$ and $\rho^{eff}$ depend on $s$ and are
in general non-local operators in space, of convolution form if the medium is statistically uniform.
A periodic medium fits this description if the exact position of any one interface is regarded
as a random variable, uniformly over one period, $(0,h]$ say. These relations are supposed to apply independently of the equation of motion: any displacement $u$ can be maintained by application
of appropriate body-force. If, however it is required only to identify some effective relations that
reproduce just a free wave travelling in the direction of positive $x$, there is only one such wave
in the present context. It has the spatial dependence $e^{\mu x}$ (times a periodic function) in any realization and it is natural and consistent to define the effective displacement to be
$e^{\mu x}$ times the mean value of the periodic function. Since $\mu$ is known as a function
of $s$, it is possible to replace the relations (3.1) by the local relations
\begin{equation}
\sigma^{eff} = \hat E^{eff}e^{eff},\;\;\;p^{eff} = \hat\rho^{eff}v^{eff},
\end{equation}
where
\begin{eqnarray}
\hat E^{eff}(s) &=& \tilde E^{eff}(s,\mu(s)) + (s/\mu(s)\tilde S_1(s,\mu(s)),\nonumber\\
\hat\rho^{eff}(s) &=& (\mu(s)/s)\tilde S_2(s,\mu(s)) + \tilde\rho^{eff}(s,\mu(s)),
\end{eqnarray}
the tilde denoting Laplace transformation with respect to $x$. The fields $\sigma^{eff}$
and $p^{eff}$ are defined in the same way as $u^{eff}$ while $e^{eff} = du^{eff}/dx$.

Note, however, that, unless $S_1=S_2=0$, a wave travelling from left to right would be described
by different $\hat E^{eff}$ and $\hat\rho^{eff}$ because $\tilde S_1$ and $\tilde S_2$
will not be even functions of $\mu$.

Henceforth, the effective relations applicable to waves travelling in the positive direction
will be described by equations (3.2) except that the hat symbols will be dropped. With this
convention, the impedance at the surface of effective medium occupying the half-space $x>0$
is $Z^{eff}\sqrt{E^{eff}\rho^{eff}}$. However, to avoid any possible confusion relating to the sign
of the square root, the alternative definition $Z^{eff} = -\sigma^{eff}/v^{eff} \equiv
-\sigma^{eff}/(s u^{eff})$ is preferred.

Before concluding this section, one additional refinement has to be mentioned. While it is important
that $\sigma^{eff}$ and $p^{eff}$ must be defined by averaging their periodic parts over an
entire period (to ensure satisfaction of the equation of motion), it is possible to define
$u^{eff}$ as a weighted average. In the most general case, this would mean averaging the periodic
part of $u$, multiplied by any periodic function $w(x)$ with period $h$ and mean value $1$. This
includes, in particular, averaging the periodic part of $u$ over any one phase, or subset
of phases. The resulting effective parameters of course depend on the weighting that is selected.

\section{Comparisons between effective and actual impedance}
\setcounter{equation}{0}

The impedance at the surface of a half-space $x>X$, composed of the layered $n$-phase
composite considered in Section 2, is $Z(X) = -\sigma(X)/su(X)$ and so, if $x_{r-1}\leq X < x_r$, 
\begin{equation}
Z(X) = -\frac{Z_r(A_r\sinh[k_r(X-x_{r-1})]+B_r\cosh[k_r(X-x_{r-1})])}
{(A_r\cosh[k_r(X-x_{r-1})]+B_r\sinh[k_r(X-x_{r-1})])},
\end{equation}
having employed the relation $k_rE_r = s Z_r$ which follows from (2.6) and (2.10). An alternative
view of the same result is that $Z(X)$ is the impedance at the surface of a half-space $x>0$,
composed of material in which interfaces in the above material
have been shifted a distance $X$ to the left. In our ``random material''
interpretation, $X$ is a random variable, uniformly distributed on $[0,h)$.

Some comparisons between actual and effective impedance are presented below, with $s = -i(2\pi f)
+ \epsilon$, where $\epsilon$ is positive (to facilitate selection of the correct eigenvalue)
but small enough to be disregarded. Thus, impedances are plotted as functions of frequency $f$.

\subsection{A two-phase laminate}

Following \cite{SNN13}, the properties of the laminate are taken as

$E_1 = 8 \times 10^9\, {\rm Pa},\;\;\rho_1 = 1180\,{\rm kg}/{\rm m}^3,\;\;h_1 = 3\,{\rm mm}$,\\
$E_2 = 300\times 10^9\,{\rm Pa},\;\;\rho_2 = 8000\,{\rm kg}/{\rm m}^3,\;\;h_2 =1.3\,{\rm mm}$.

This is the material whose dispersion relation is illustrated in Figs. 1 and 2. Figure 3 and 4
show plots of impedance at front and at the centre of each phase. 
\begin{figure}
\centering
\includegraphics[scale=0.35]{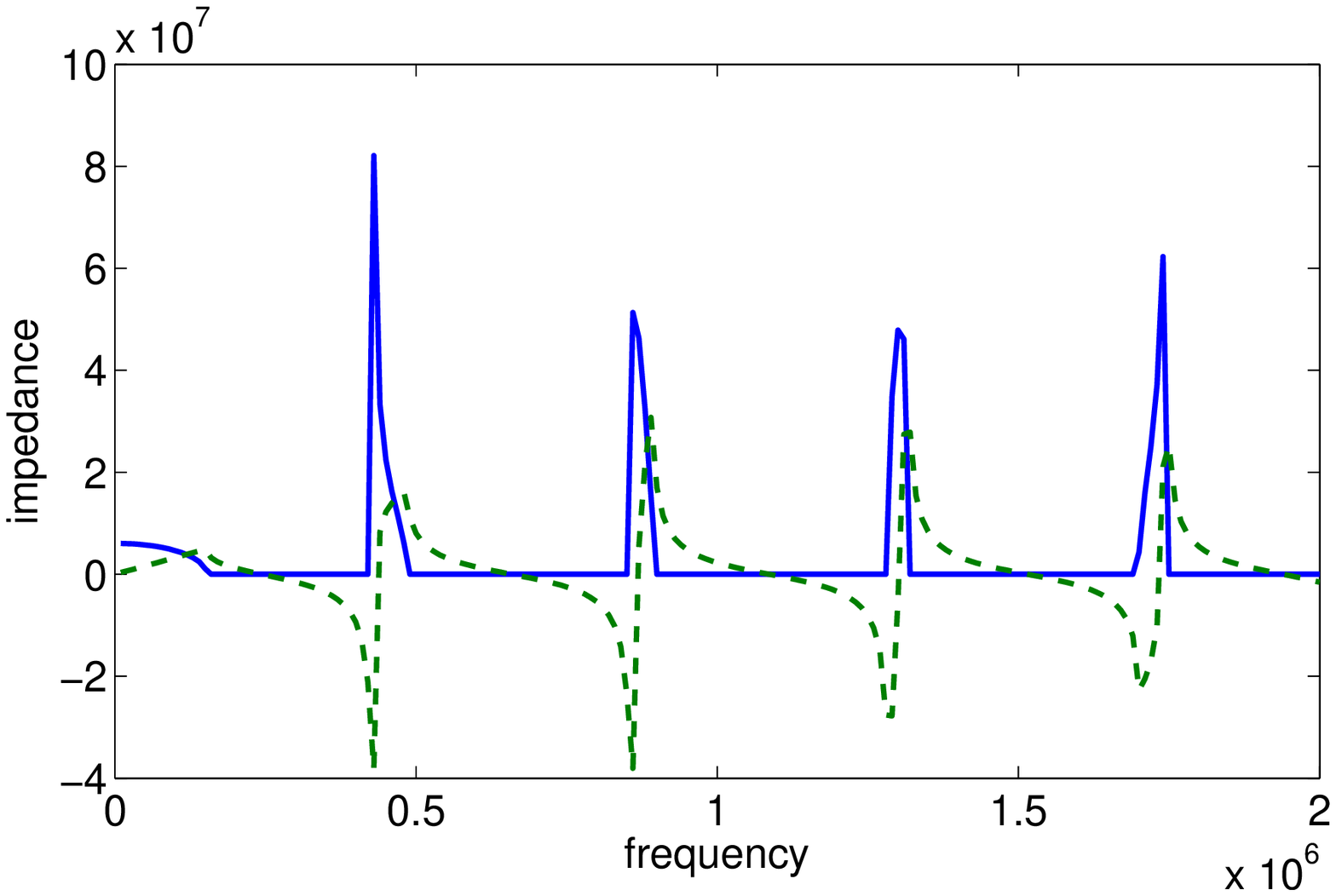}
\includegraphics[scale=0.35]{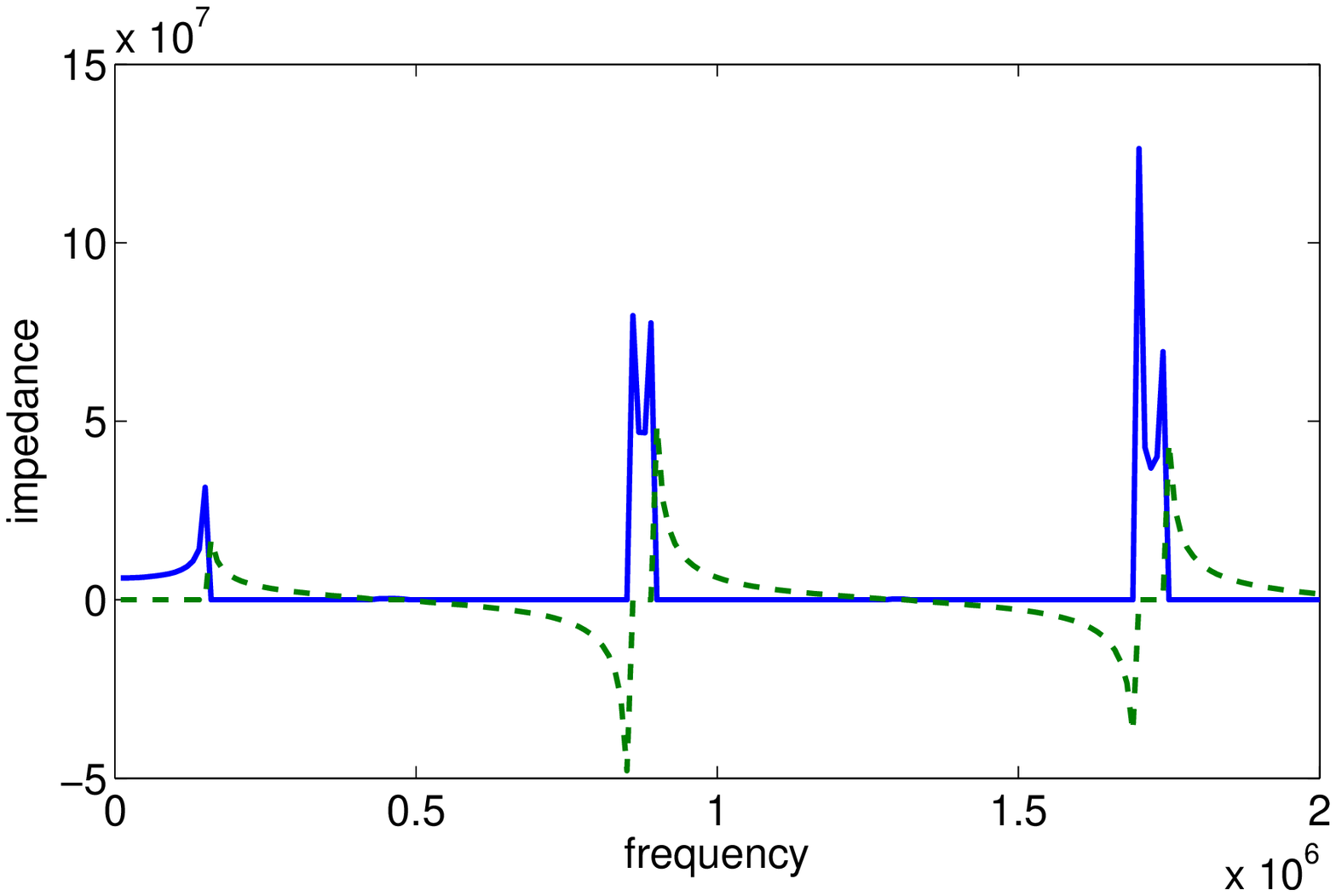}\\
(a)\hskip 6cm (b)
\caption{\small Impedance versus frequency (a) at front of phase 1, (b) mid-phase 1.}
\end{figure}
\begin{figure}
\centering
\includegraphics[scale=0.35]{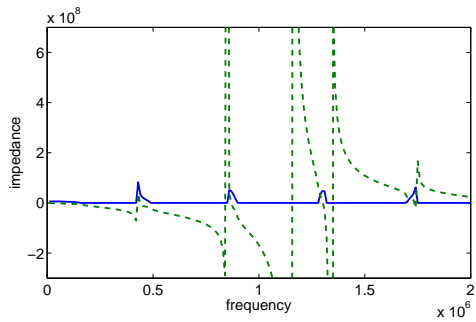}
\includegraphics[scale=0.35]{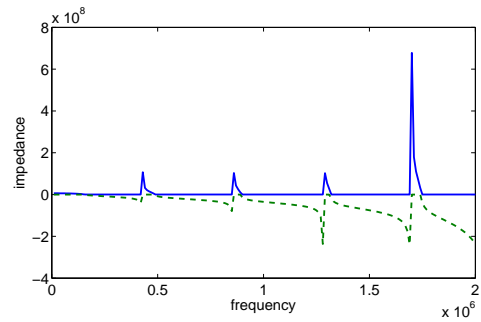}\\
(a)\hskip 6cm (b)
\caption{\small Impedance versus frequency (a) at front of phase 2, (b) mid-phase 2.}
\end{figure}
The differences in the impedances at different points are pronounced. Note, however, that
all are pure imaginary in the stop bands, and in the pass bands all have positive real part.
This corresponds to positive mean rate of working by the pressure at the surface.

Since any ``effective'' impedance will reflect some kind of averaged response, it is evident
that it cannot approximate the actual response, except at low frequencies. To facilitate
study of this, actual impedances are first plotted over a more restricted frequency range,
in Figs. 5 and 6. Then, Fig. 7 shows the effective impedance, calculated employing the ``unweighted''
average for effective displacement.
\begin{figure}
\centering
\includegraphics[scale=0.35]{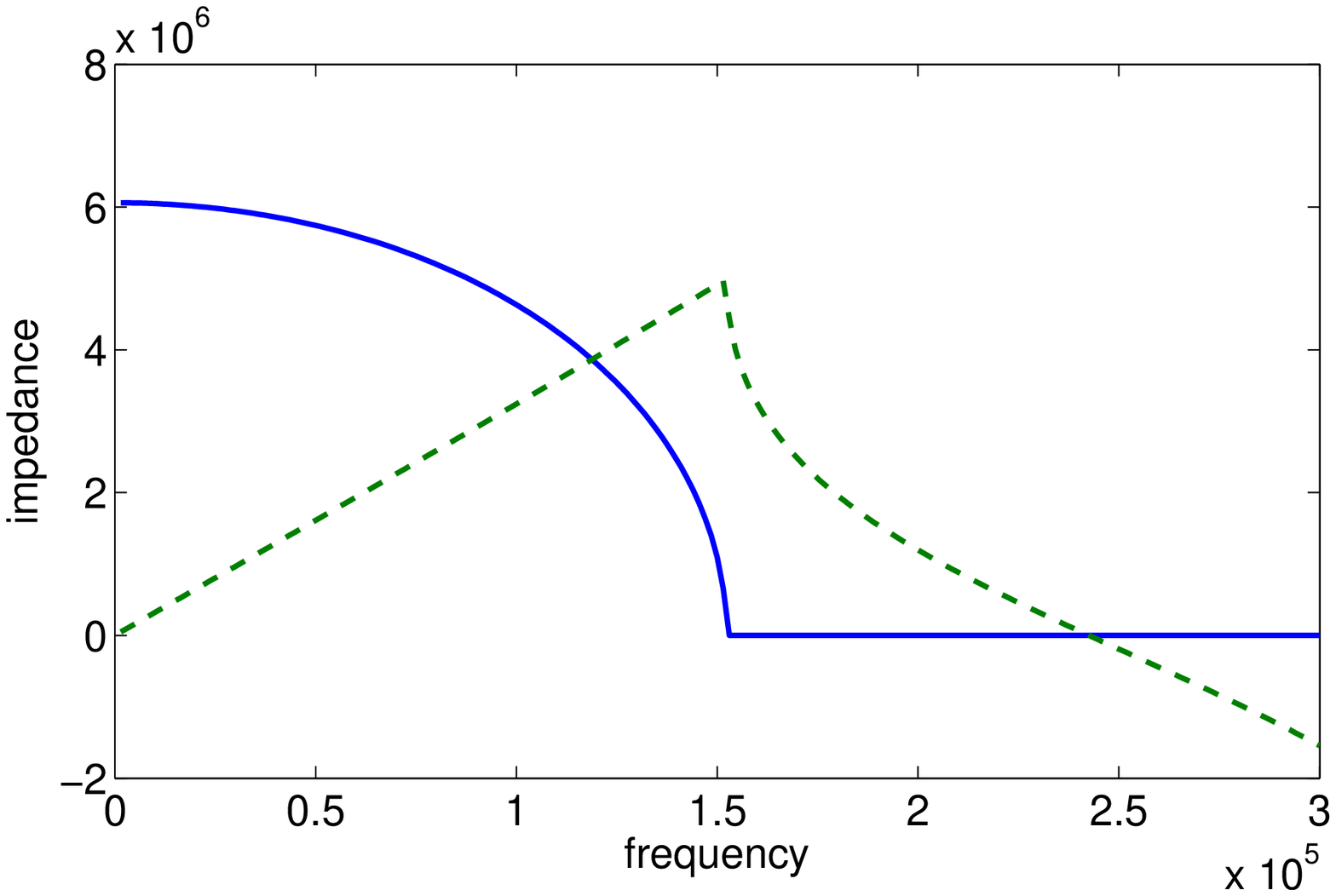}
\includegraphics[scale=0.35]{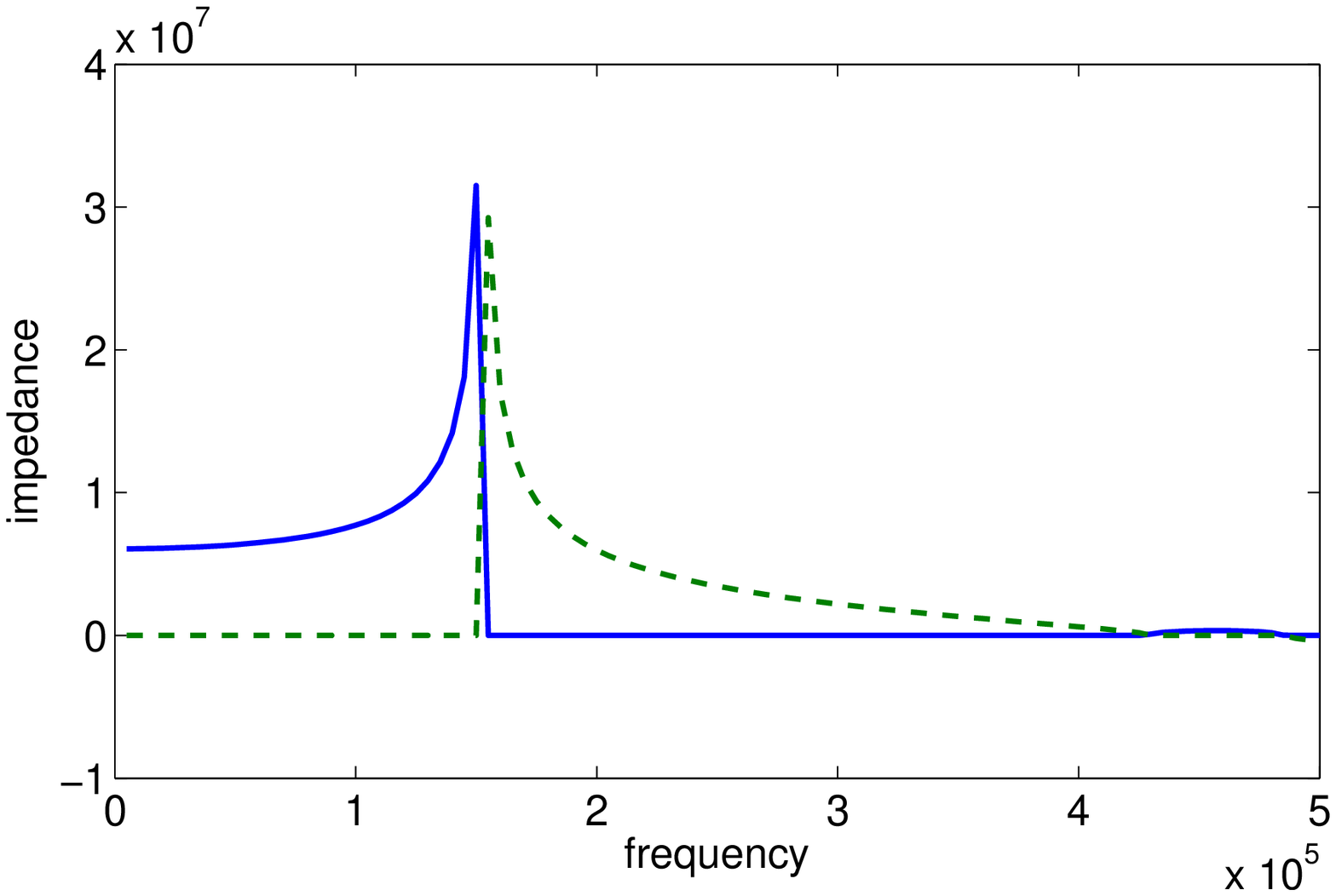}\\
(a)\hskip 6cm (b)
\caption{\small Impedance versus frequency (a) at front of phase 1, (b) mid-phase 1.
Smaller frequency range.}
\end{figure}
\begin{figure}
\centering
\includegraphics[scale=0.35]{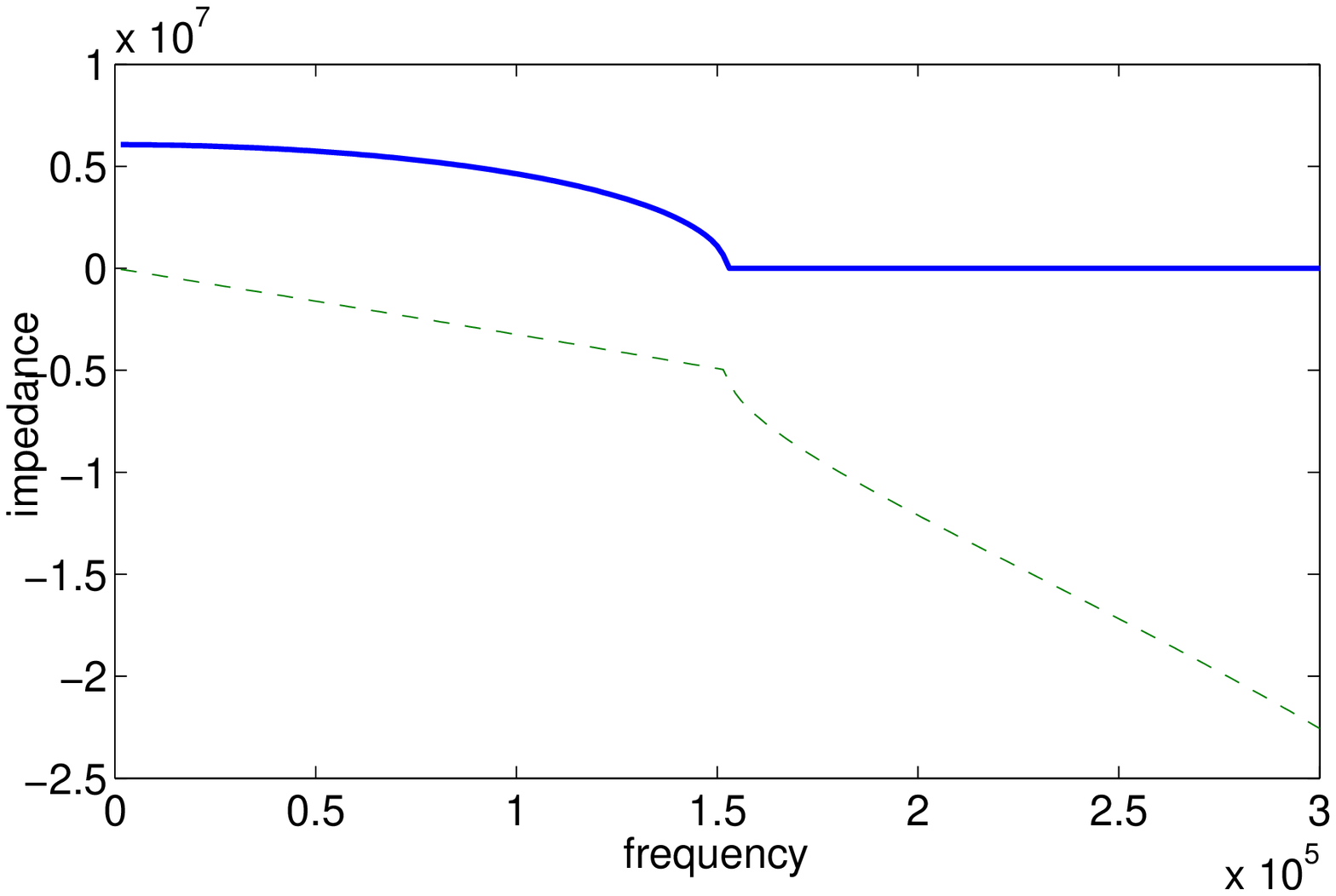}
\includegraphics[scale=0.35]{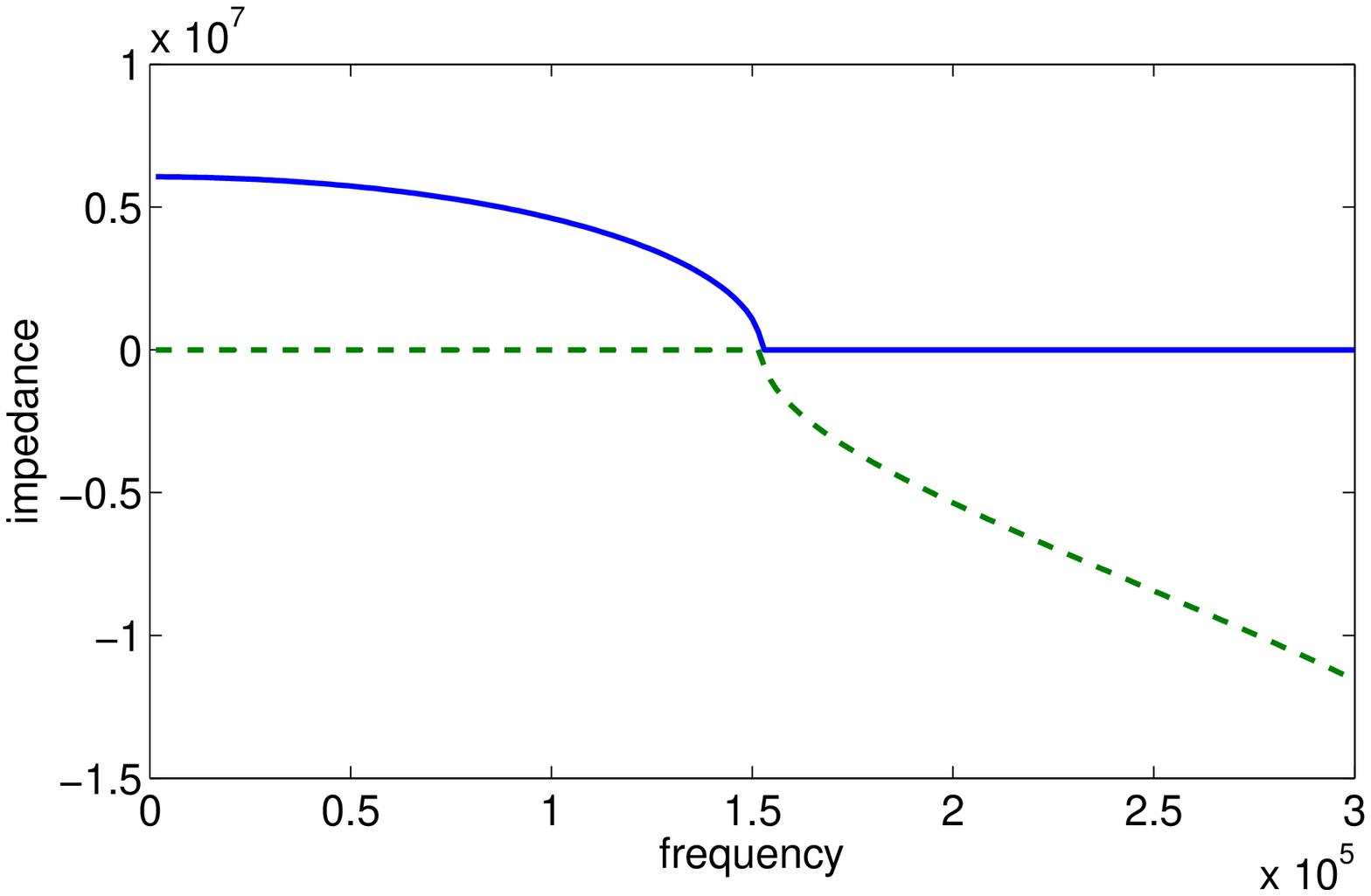}\\
(a)\hskip 6cm (b)
\caption{\small Impedance versus frequency (a) at front of phase 2, (b) mid-phase 2.
Smaller frequency range.}
\end{figure}

\begin{figure}
\centering
\includegraphics[scale=0.5]{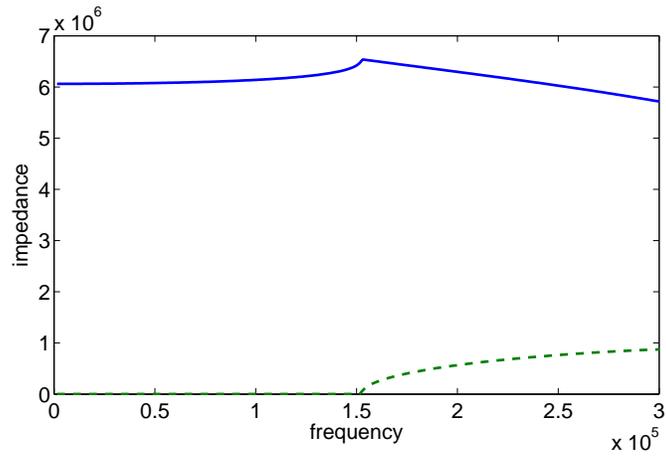}
\caption{\small ``Unweighted'' effective impedance.}
\end{figure}
Figure 8 shows the corresponding ``unweighted'' effective modulus and effective density. These values
confirm that $Z^{eff}$ in Fig. 7 is equal to $\sqrt{(E^{eff}\rho^{eff})}$.
\begin{figure}
\centering
\includegraphics[scale=0.35]{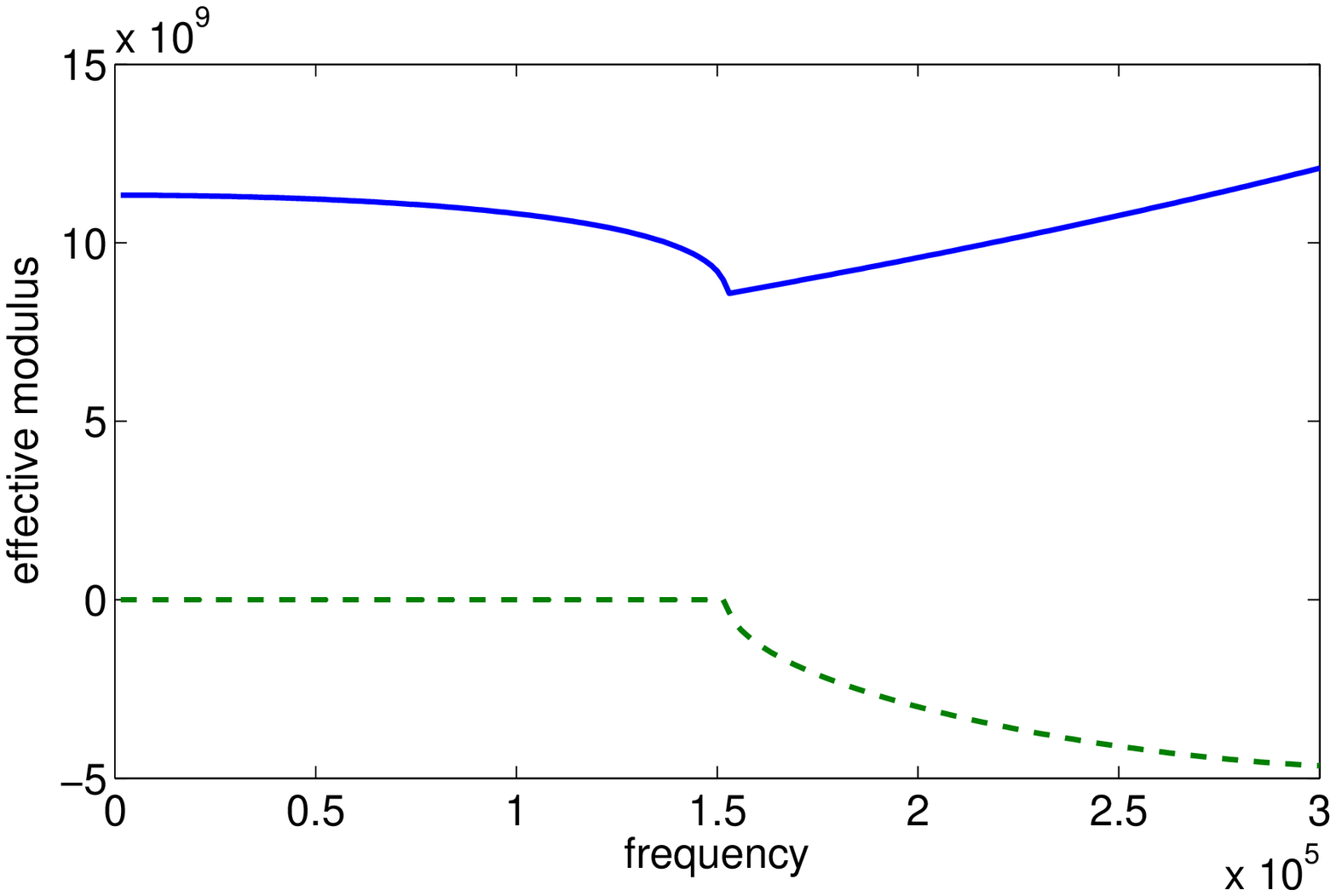}
\includegraphics[scale=0.35]{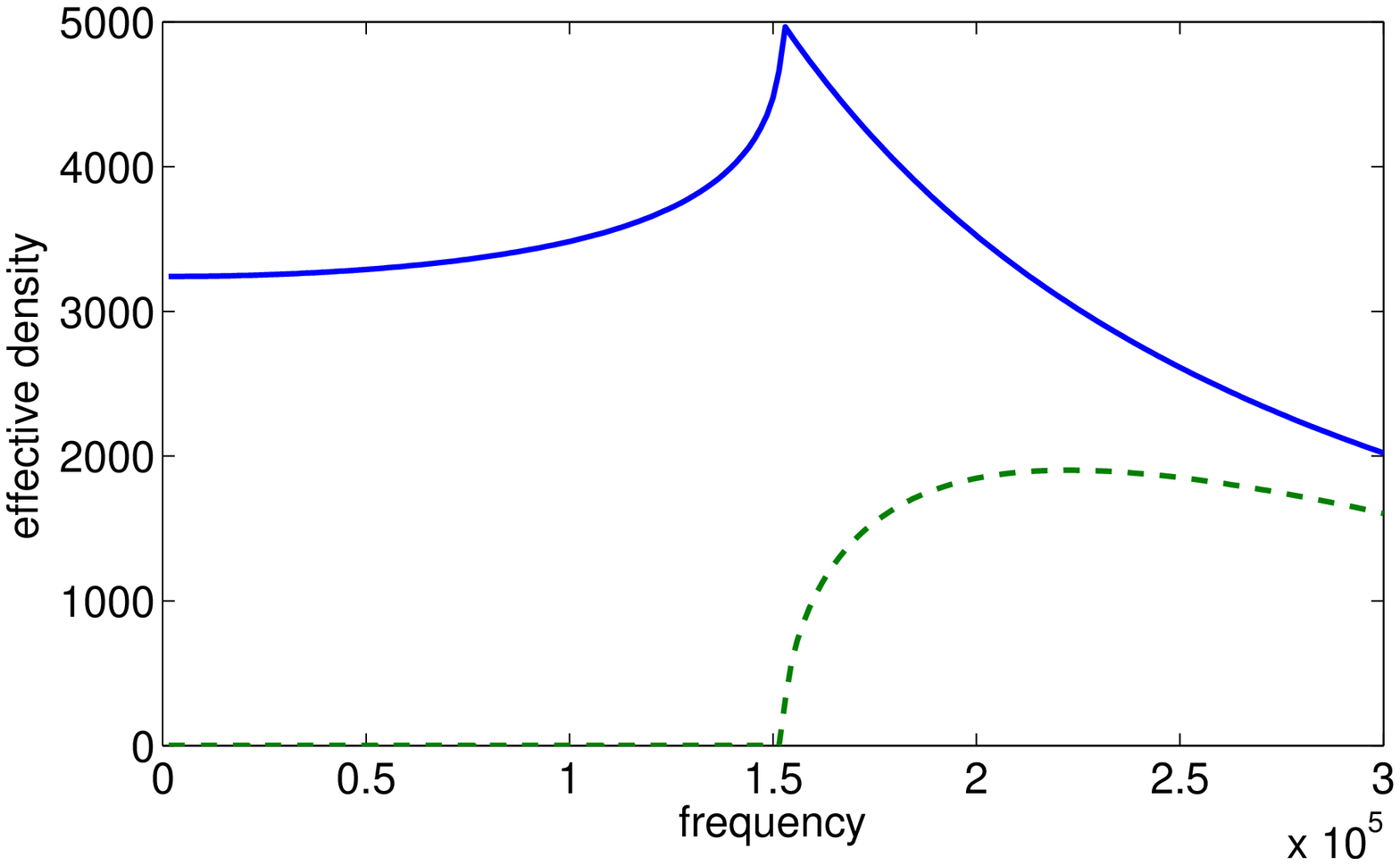}\\
(a)\hskip 6cm (b)
\caption{\small (a) effective modulus, (b) effective density, both unweighted.}
\end{figure}
It is inevitable that the unweighted $Z^{eff}$ cannot reproduce the actual impedance, because this varies from point to point. It is also not to be expected that $Z^{eff}$ gives the exact mean value of $Z$,
because the mean value of a fraction is now equal to the fraction of the corresponding mean values.

Figure 9 shows plots of effective impedance, weighted on phase 1 (Fig. 9a), and on phase 2 (Fig. 9b).
Figure 9a should be compared with Fig. 5b, and Fig. 9b should be compared with Fig. 6b. Reasonable
agreement can be recognised for frequencies up to about 100 kHz -- an appreciable range, but still
within the first pass band.

It should be noted that the real parts of effective impedance, weighted or not, are qualitatively wrong
(i.e. they are not zero) in the first stop band. This is in part associated with the ambiguity
in the definition of $\mu h$. In the figures plotted so far, its phase was taken as $+\pi$,
consistent with Fig. 2. If, however, the phase is taken instead as $-\pi$, the resulting values
for the weighted effective impedances are shown in Fig. 10. The corresponding plots
in Figs. 9 and 10 display exactly the same values in the pass band but differ in the stop band,
where the phases of $\mu h$ are different, and hence the effective fields are calculated as
the mean values of different periodic functions.
\begin{figure}
\centering
\includegraphics[scale=0.35]{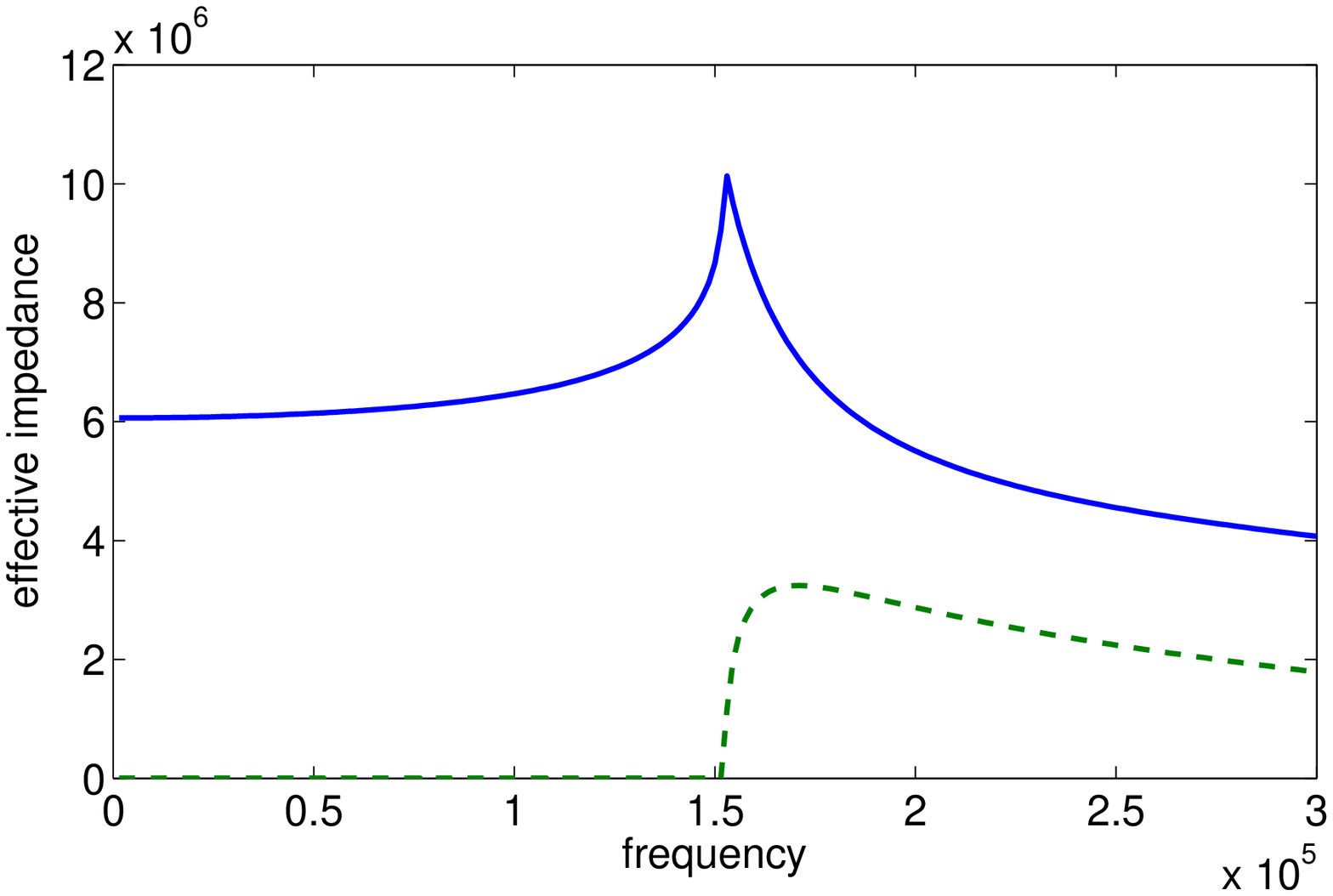}
\includegraphics[scale=0.35]{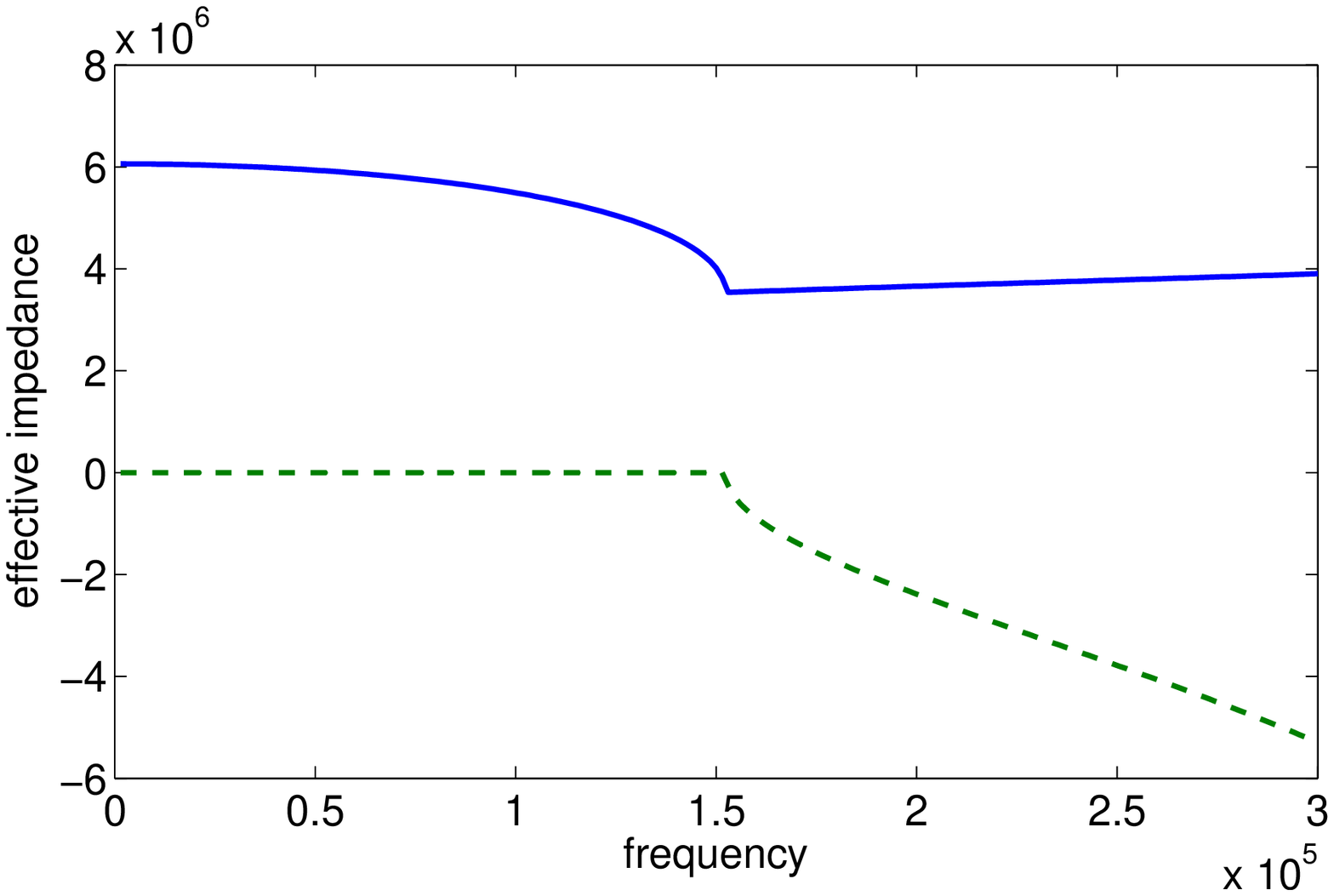}\\
(a)\hskip 6cm (b)
\caption{\small Effective impedance (a) phase 1 average, (b) phase 2 average.}
\end{figure}
\begin{figure}
\centering
\includegraphics[scale=0.35]{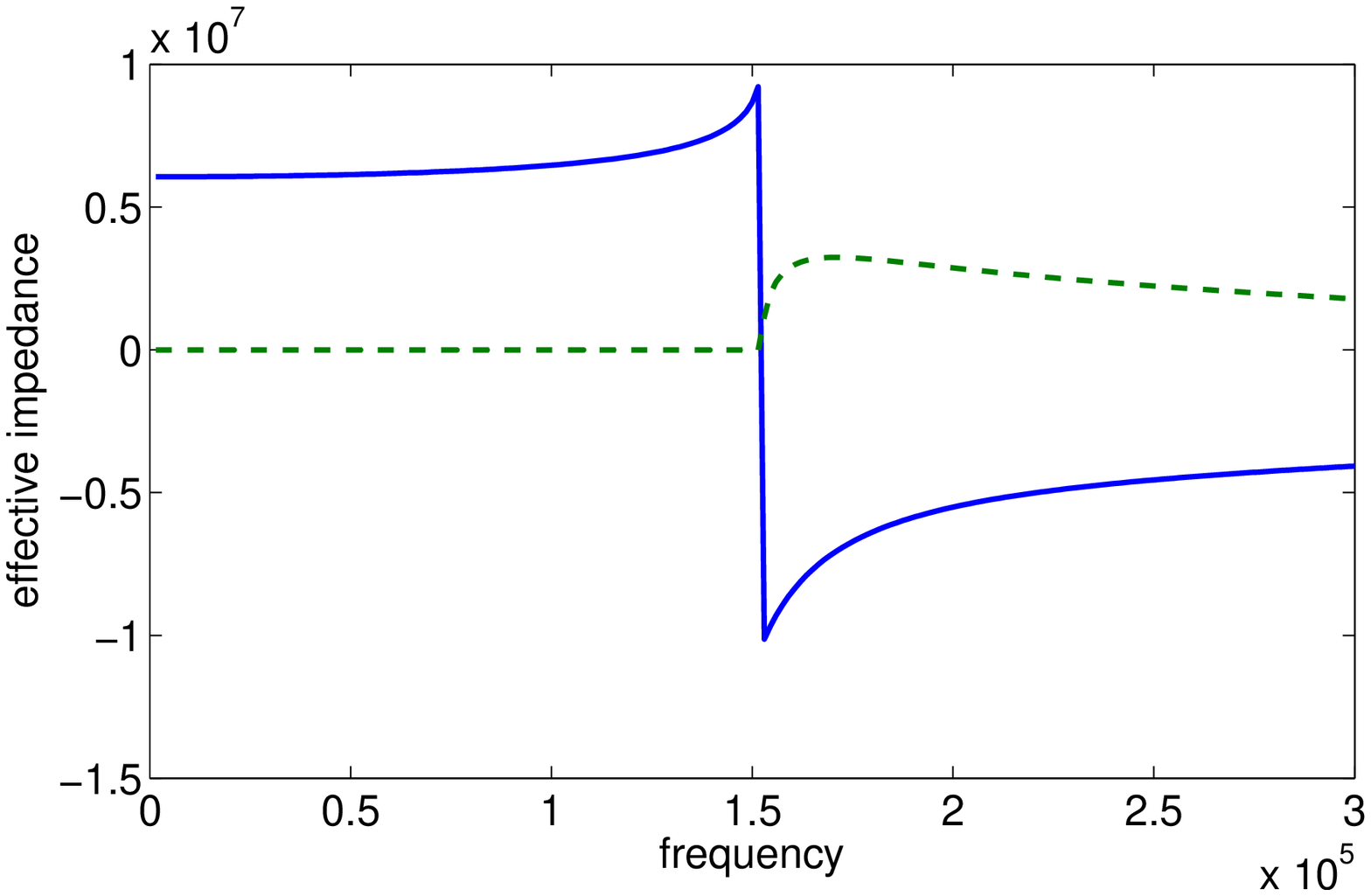}
\includegraphics[scale=0.35]{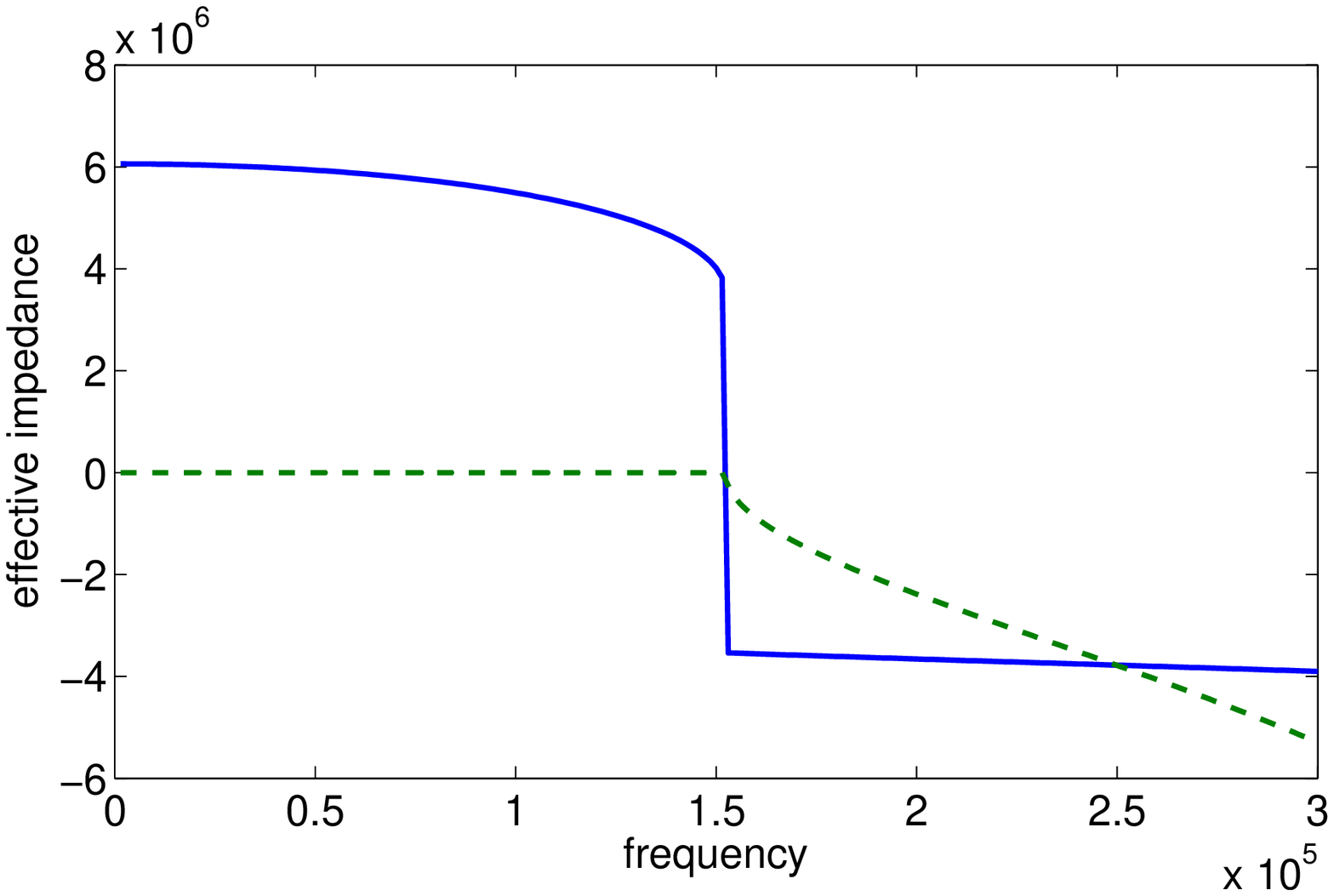}\\
(a)\hskip 6cm (b)
\caption{\small Effective impedance (a) phase 1 average, (b) phase 2 average, with ``$-\pi$''
phase for $\mu h$ in the stop band.}
\end{figure}

\subsection{A three-material laminate}

Some results are now presented for a laminate composed of three different materials, which was also
studied in \cite{SNN13}. It is convenient to describe it, however, as a
``five-phase'' periodic laminate, with properties as follows.

$E_1 = 8\times 10^9\,{\rm Pa},\;\;\rho_1 = 1180\,{\rm kg}/{\rm m}m^3,\;\;h_1 = 1.45\,{\rm mm}$,\\
$E_2 = 0.02\times 10^9\,{\rm Pa},\;\;\rho_2 = 1100\,{\rm kg}/{\rm m}m^3,\;\;h_2 = 0.5\,{\rm mm}$,\\
$E_3 = 300\times 10^9\,{\rm Pa},\;\;\rho_3 = 8000\,{\rm kg}/{\rm m}m^3,\;\;h_3 = 0.4\,{\rm mm}$,\\ 
$E_4 = 0.02\times 10^9\,{\rm Pa},\;\;\rho_4 = 1100\,{\rm kg}/{\rm m}m^3,\;\;h_4 = 0.5\,{\rm mm}$,\\
$E_5 = 8\times 10^9\,{\rm Pa},\;\;\rho_5 = 1180\,{\rm kg}/{\rm m}m^3,\;\;h_5 = 1.45\,{\rm mm}$.

It could equally well be described as a ``four-phase'' material but the specification given
above shows the symmetry and was in fact used in the computations. The interesting feature
of this composite is that the dense ``middle'' phase 3 is surrounded by soft phase 2 material
and produces a resonance at a relatively low frequency. A plot of the normalized dispersion relation,
$\mu h$ versus frequency $f$, is shown in Fig. 11a. There is, in fact, a small stop band around
$f=23\,$kHz. This is shown in the more detailed plot of Fig. 11b.

Effective response is subject to the limitations already exposed in the two-phase case, so
the point will not be laboured again. Figure 12 shows the actual impedance at the leading edge of
``phase 1'', which corresponds to the middle of ``material 1'', for the extended
frequency range. Figure 13a shows the detail, for frequencies around the first stop band,
while Fig. 13b shows the effective impedance in the same frequency range,
calculated using the corresponding weighted average of the displacement. Note, however, that
the first stop band is narrow (from about 23 to 23.7 kHz) and that the effective impedance
in the second pass band remains reasonably representative of, though somewhat larger
than, the actual response.

\begin{figure}
\centering
\includegraphics[scale=0.35]{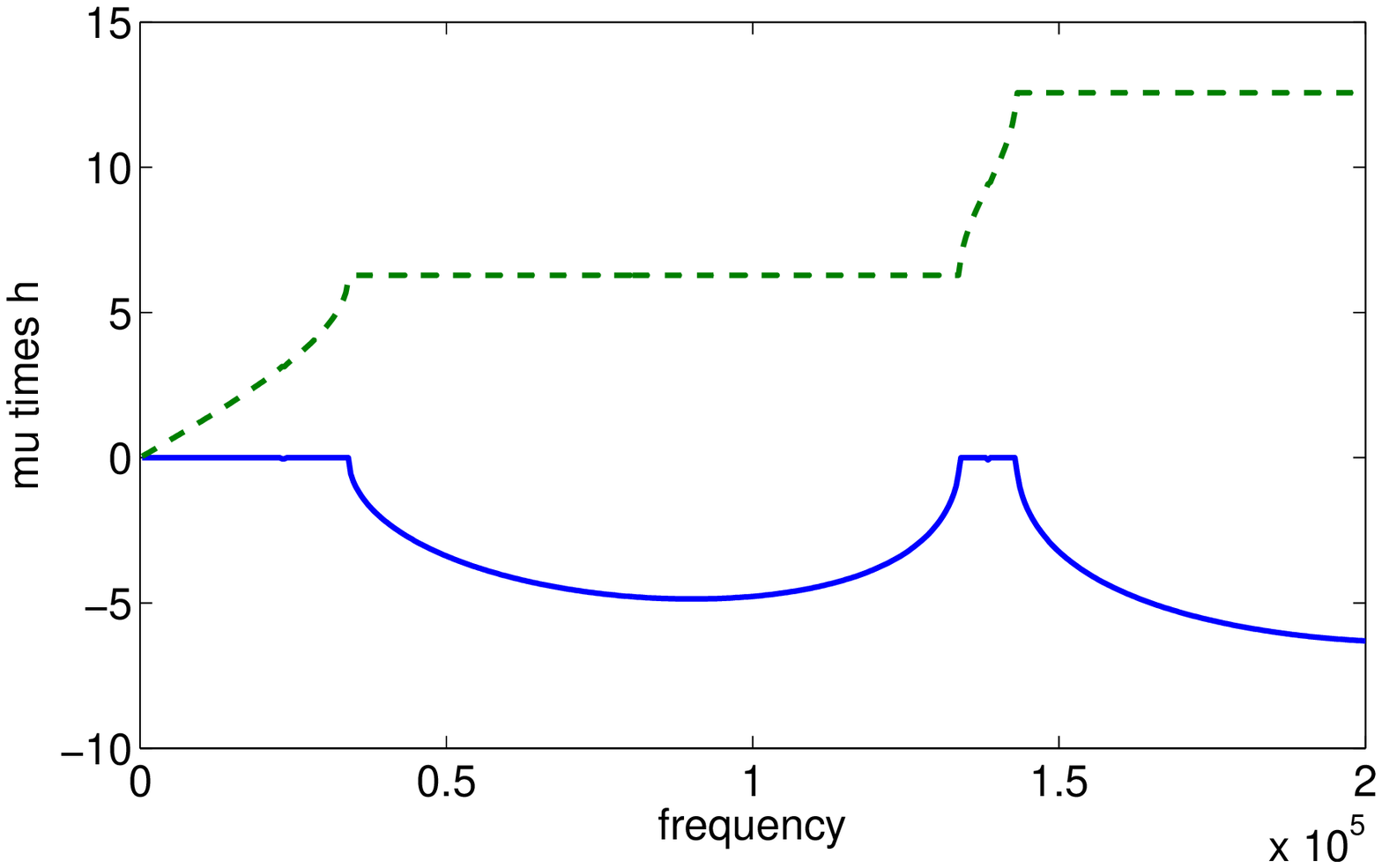}
\includegraphics[scale=0.35]{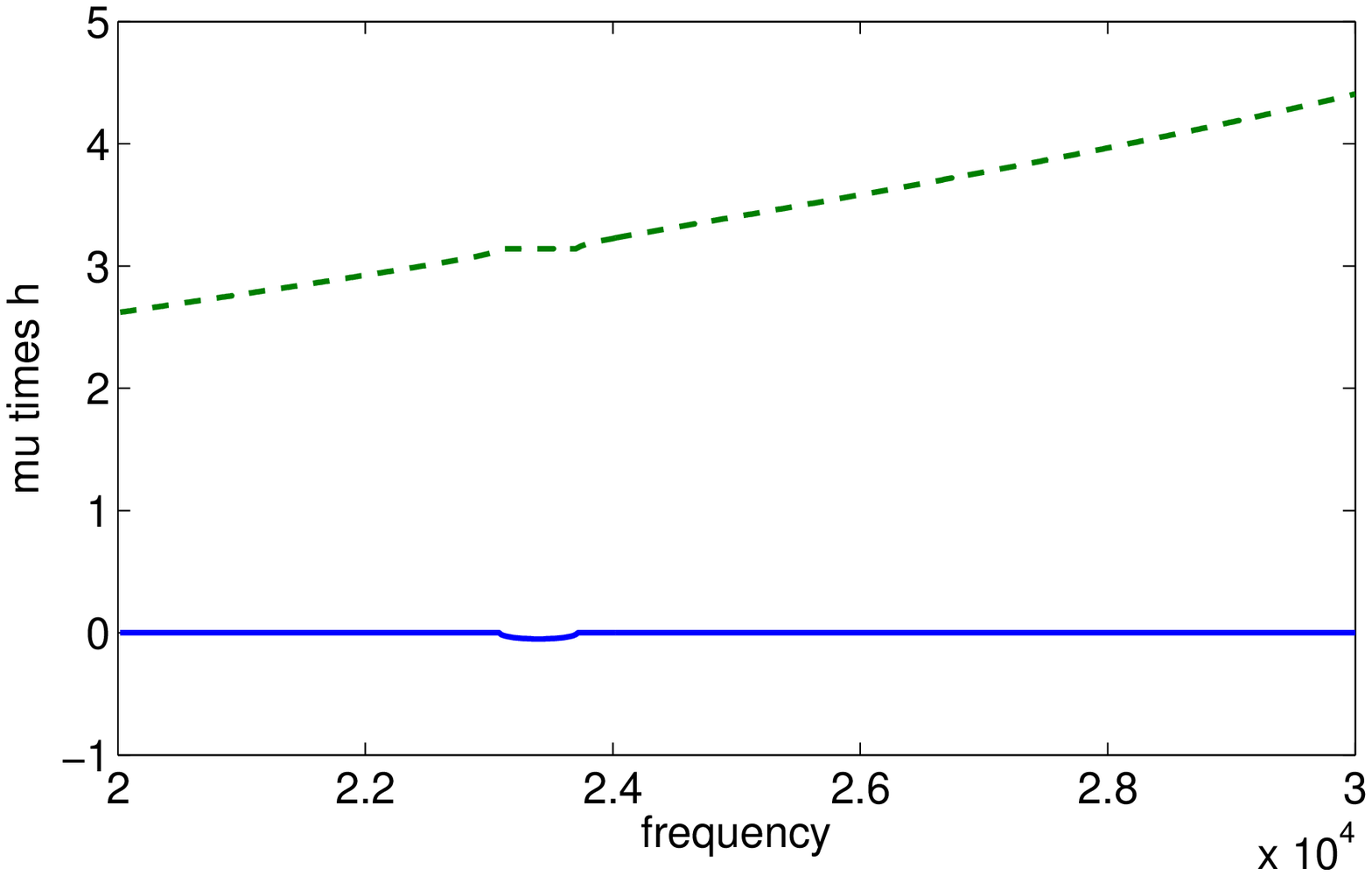}\\
(a)\hskip 6cm (b)
\caption{\small Normalized dispersion relation for three-phase material (a) extended frequency
range, (b) detail showing first stop band.}
\end{figure}
\begin{figure}
\centering
\includegraphics[scale=0.5]{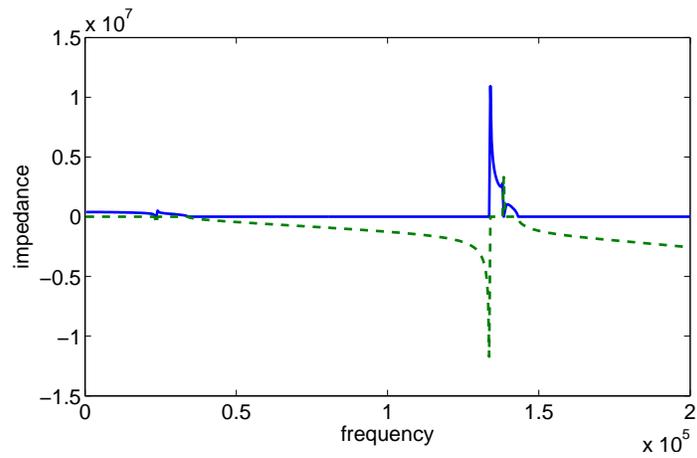}
\caption{\small Actual impedance mid-material 1.}
\end{figure}
\begin{figure}
\centering
\includegraphics[scale=0.35]{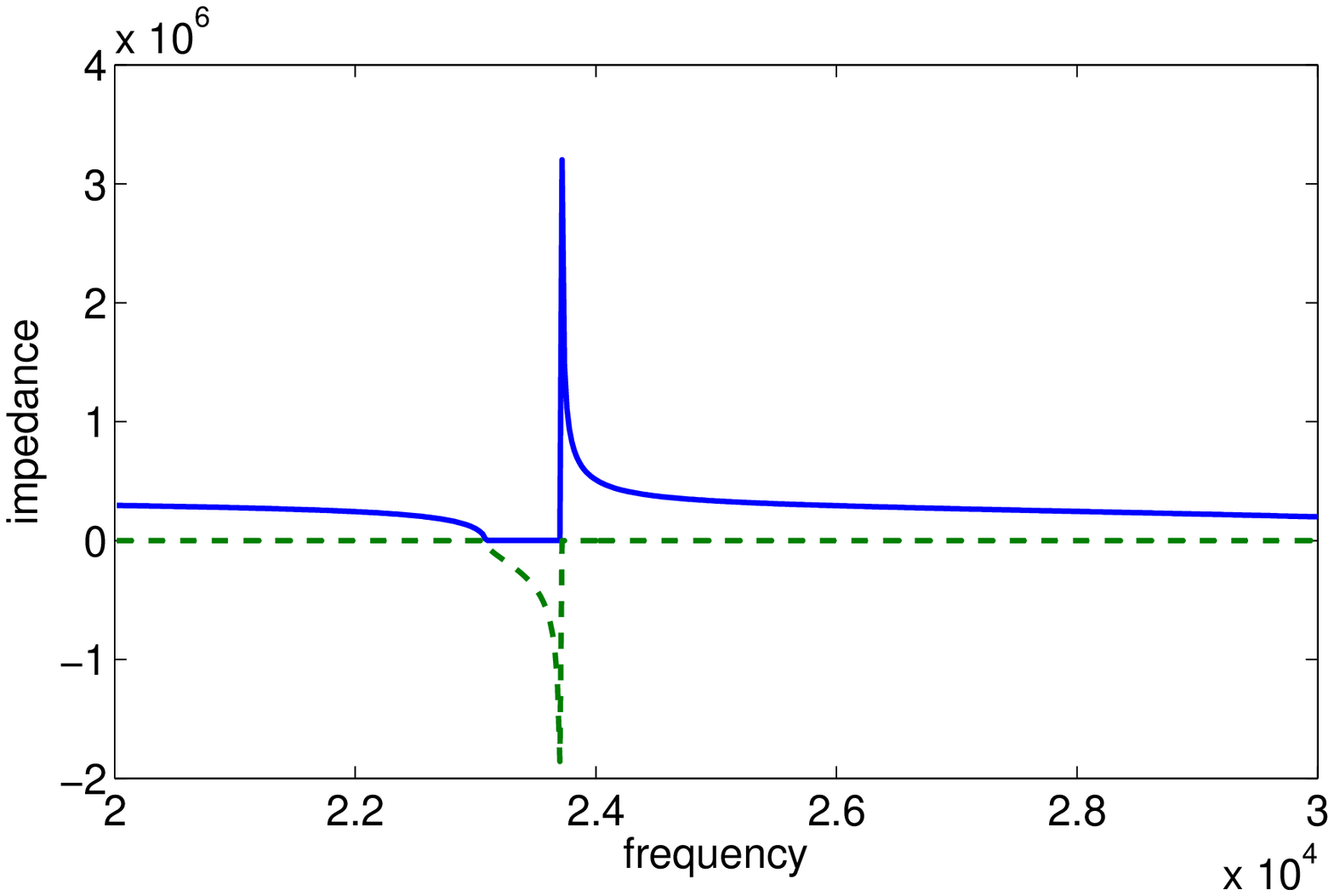}
\includegraphics[scale=0.35]{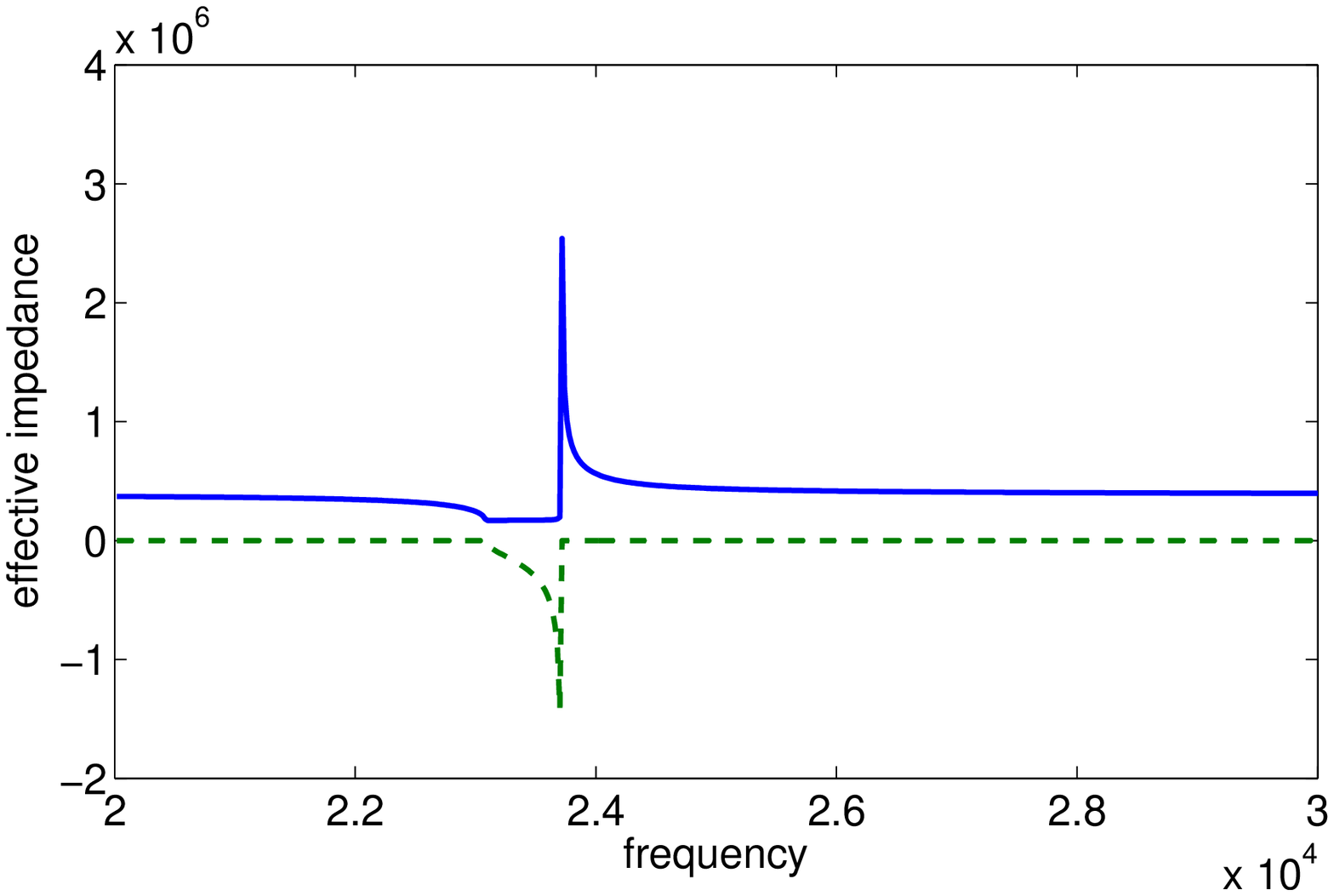}\\
(a)\hskip 6cm (b)
\caption{\small (a) Actual impedance mid-material 1, (b) corresponding effective impedance.}
\end{figure}

\section{An alternative view}
\setcounter{equation}{0}

What has been described so far is based on what could be regarded as the ``conventional'' view of
effective relations, for composites with periodic microstructure. Regardless of the utility or
otherwise of effective properties, it is disturbing -- at least to me -- that taking averages
of fields, each one of which displays exactly the same stop band structure, should result in
an ``effective impedance'' which depends on which branch of the (exact) dispersion relation
should be selected, and which fails altogether to display even the first stop band. It is worth
recalling that the exact definition of effective constitutive relations, involving non-local
operators, is {\it completely independent} of the selection of any particular branch of the
dispersion relation: these relations are obtained from taking ensemble averages of expressions
involving the exact Green's function. The formulae given in \cite{JRW09}, for example, show
that these ensemble averages contain summations over all equivalent wavenumbers and so require
no judgement as to which branch is ``relevant''.

Following on from this, it seems worth contemplating the ensemble average of the Green's function
itself, $\langle G\rangle$ (or, more generally, the weighted average $\langle wGw\rangle$
introduced in \cite{JRW09}). In the one-dimensional setting considered here, the point force
at $x=0$ generates a single wave in the half-space $x <0$ which propagates from right to left
(and, if $s$ has positive real part, decays exponentially as $x\to -\infty$), and a wave in
the half-space $x>0$ which propagates from left to right and decays as $x\to +\infty$. A reasonable
definition of effective impedance of the half-space $x>X$, with $X>0$, is therefore to take
\begin{equation}
Z^{eff}(X) = -\Sigma^{eff}(X)/sG^{eff}(X)
\end{equation}
where, for a general weight $w$, $G^{eff}(X) = \langle w(X)G(X)w(0)\rangle$ while $\Sigma^{eff}(X)
= \langle \Sigma(X)\rangle $, $\Sigma(X)$ being the stress associated with the displacement $G(X)w(0)$.
Explicit expressions for both of these are given in \cite{JRW09}.

The impedance $Z^{eff}(X)$ so defined is, by its construction, independent of any choice of branch
for the dispersion relation. The penalty, however, is that it does depend on $X$: it is a periodic function
with the period of the microstructure. The results of some sample calculations follow.

\subsection{Two-phase laminate}

Figure 14 shows the effective impedance $Z^{eff}(X)$, evaluated (a) as $X\to 0$ and
(b) at $X=h/2$. These plots should be compared with
the exact impedances plotted in Figs. 5 and 6. It is already evident from Figs. 5 and 6 that
no single effective impedance can be applicable over the whole of the frequency range that
is shown. Figure 15 gives a plot $Z^{eff}(X)$ versus $X/h$ (a) when frequency $f= 100$\,kHz and
(b) when $f=150$\,kHz. The relatively small variation of $Z^{eff}$ at $100$\,kHz is consistent with
the applicability of effective medium theory at that frequency, whereas Fig. 15(b) provides evidence
that effective medium theory is not useful at $150$\,kHz. However, at least the stop band is
correctly identified, with the present definition of $Z^{eff}$.
\begin{figure}
\centering
\includegraphics[scale=0.35]{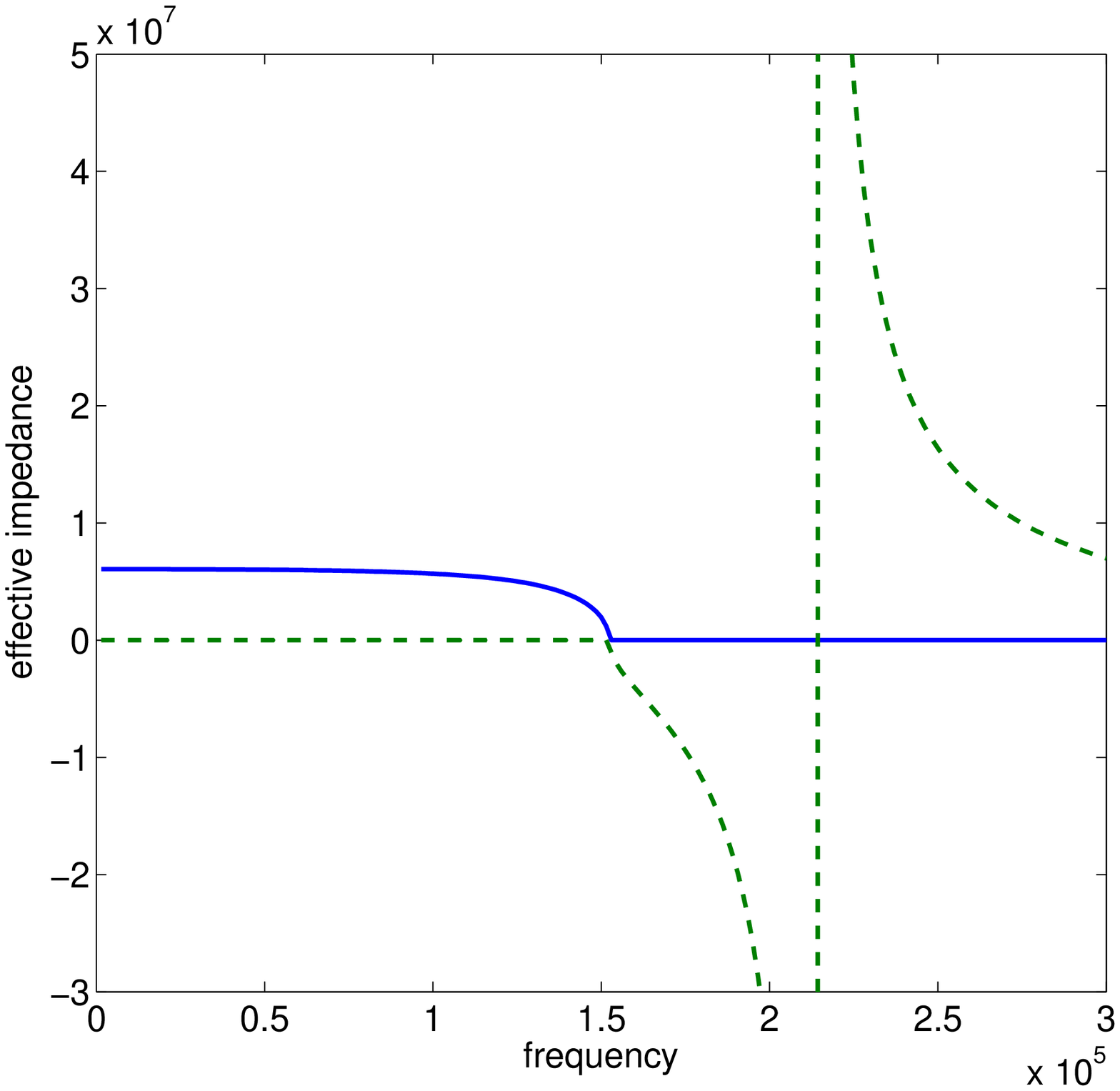}
\includegraphics[scale=0.35]{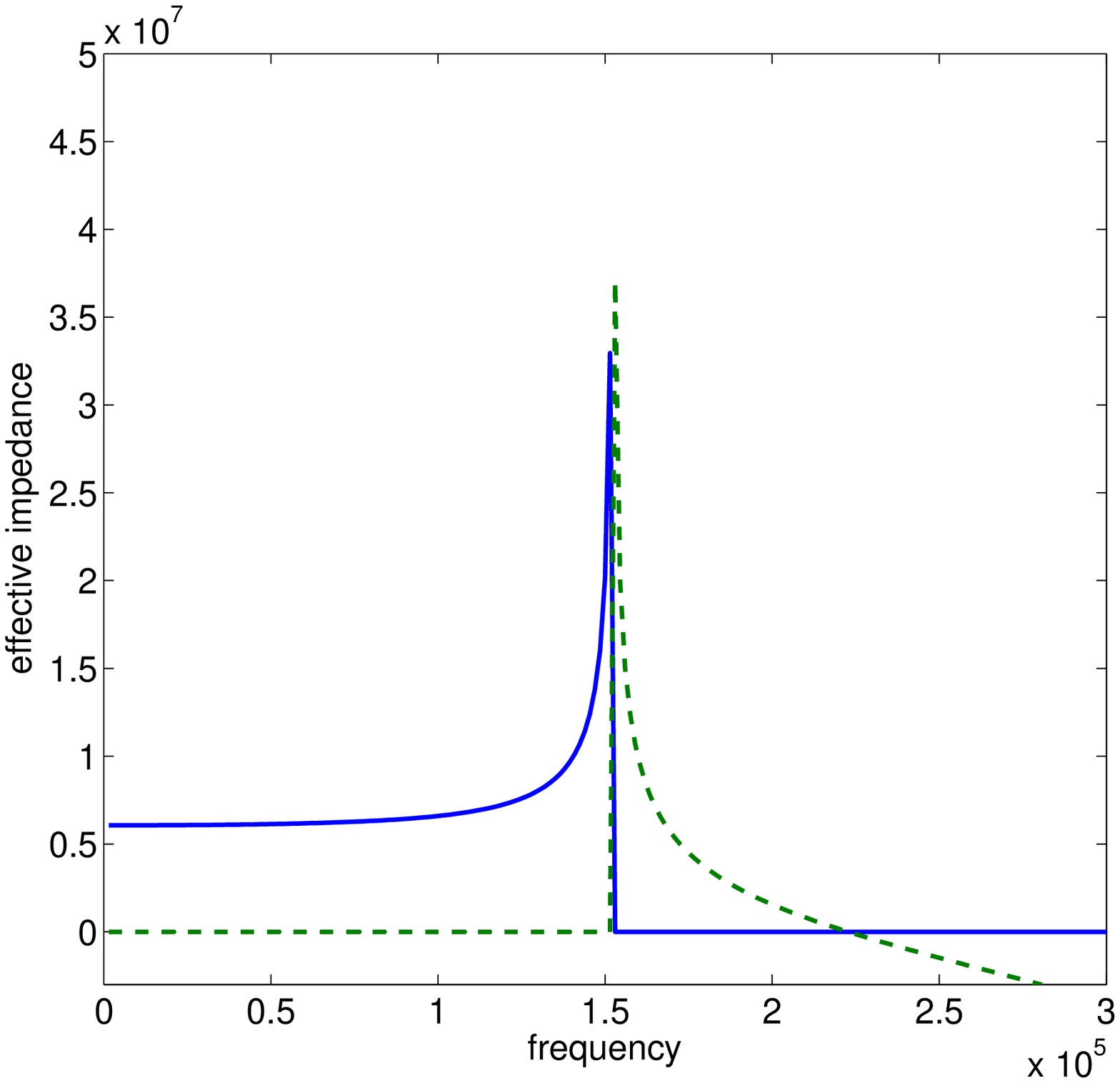}\\
(a)\hskip 6cm (b)
\caption{\small (a) Effective impedance at $X=0$, (b) effective impedance at $X=h/2$.}
\end{figure}
\begin{figure}
\centering
\includegraphics[scale=0.35]{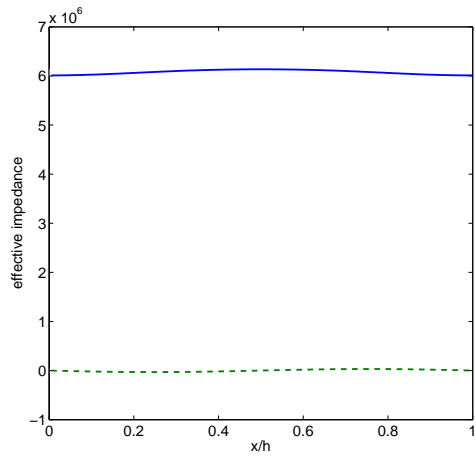}
\includegraphics[scale=0.35]{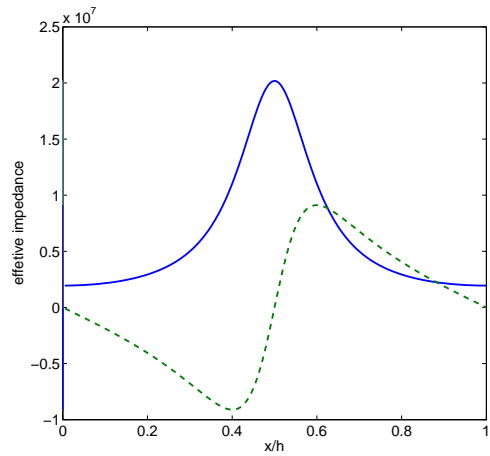}\\
(a)\hskip 6cm (b)
\caption{\small Effective impedance at $X=0$ (a) at frequency $f=100$\,kHz, (b) at $f=150$\,kHz.}
\end{figure}

\subsection{Three-phase laminate}
\begin{figure}
\centering
\includegraphics[scale=0.35]{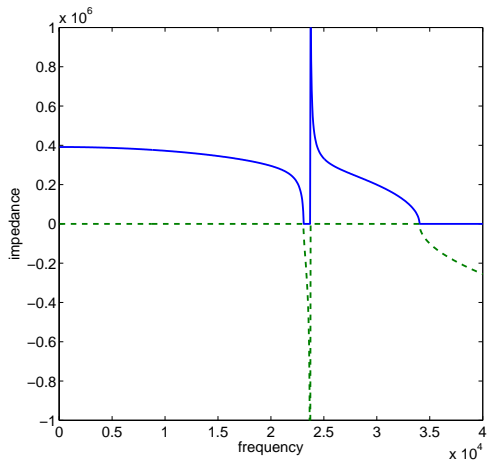}
\includegraphics[scale=0.35]{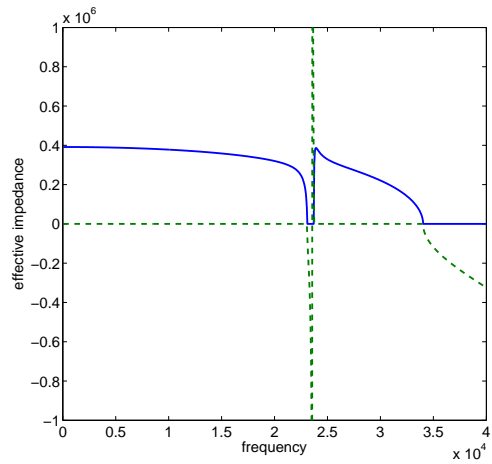}\\
(a)\hskip 6cm (b)
\caption{\small (a) Exact impedance, mid-``material 1'' for the three-phase material,
(b) effective impedance at $X=0$. }
\end{figure}

Figure 16 shows plots (a) of the exact impedance of the three-material laminate introduced in Section
4.2, with boundary $x=0$ coinciding with the front of the cell described as  ``phase 1'', and (b) the
effective impedance $Z^{eff}(0)$. The close similarity of the two plots is noteworthy. They would
differ at frequencies somewhat higher than shown but nevertheless the effective impedance captures
the first stop band exactly and provides a better estimate of the impedance in the second pass
band than the estimate reported in Section 4.2 -- compare Figs. 13 and 16. 

Figure 17 shows plots of $Z^{eff}(X)$ against $X/h$, (a) for frequency $f = 17.5$\,kHz and (b) $f = 325$\,kHz. These frequencies lie on either side of the first stop band. They demonstrate, even though
the plots of Fig. 16 display reasonable agreement, that use of effective medium theory involves some loss
of precision. 
\begin{figure}
\centering
\includegraphics[scale=0.35]{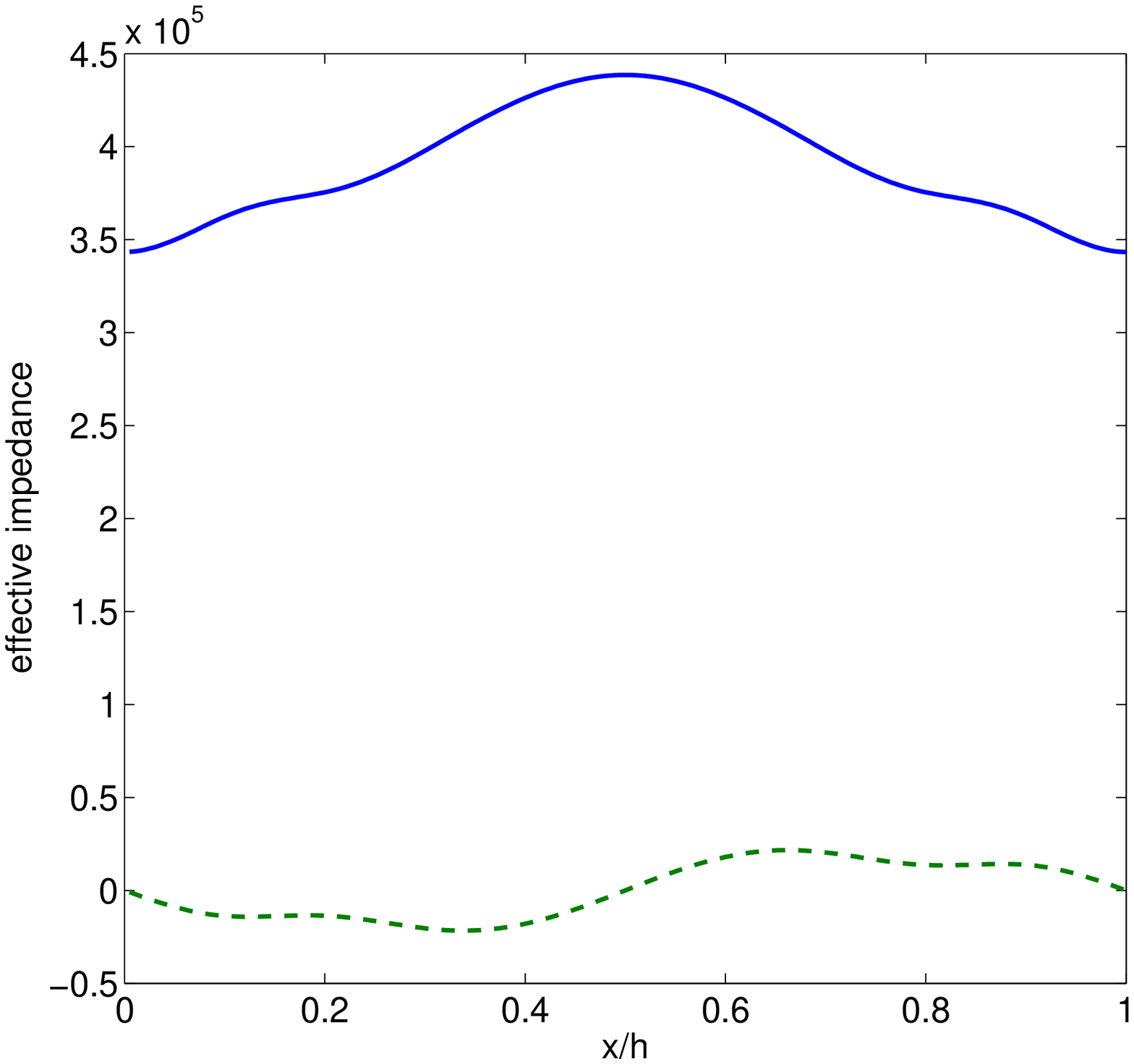}
\includegraphics[scale=0.35]{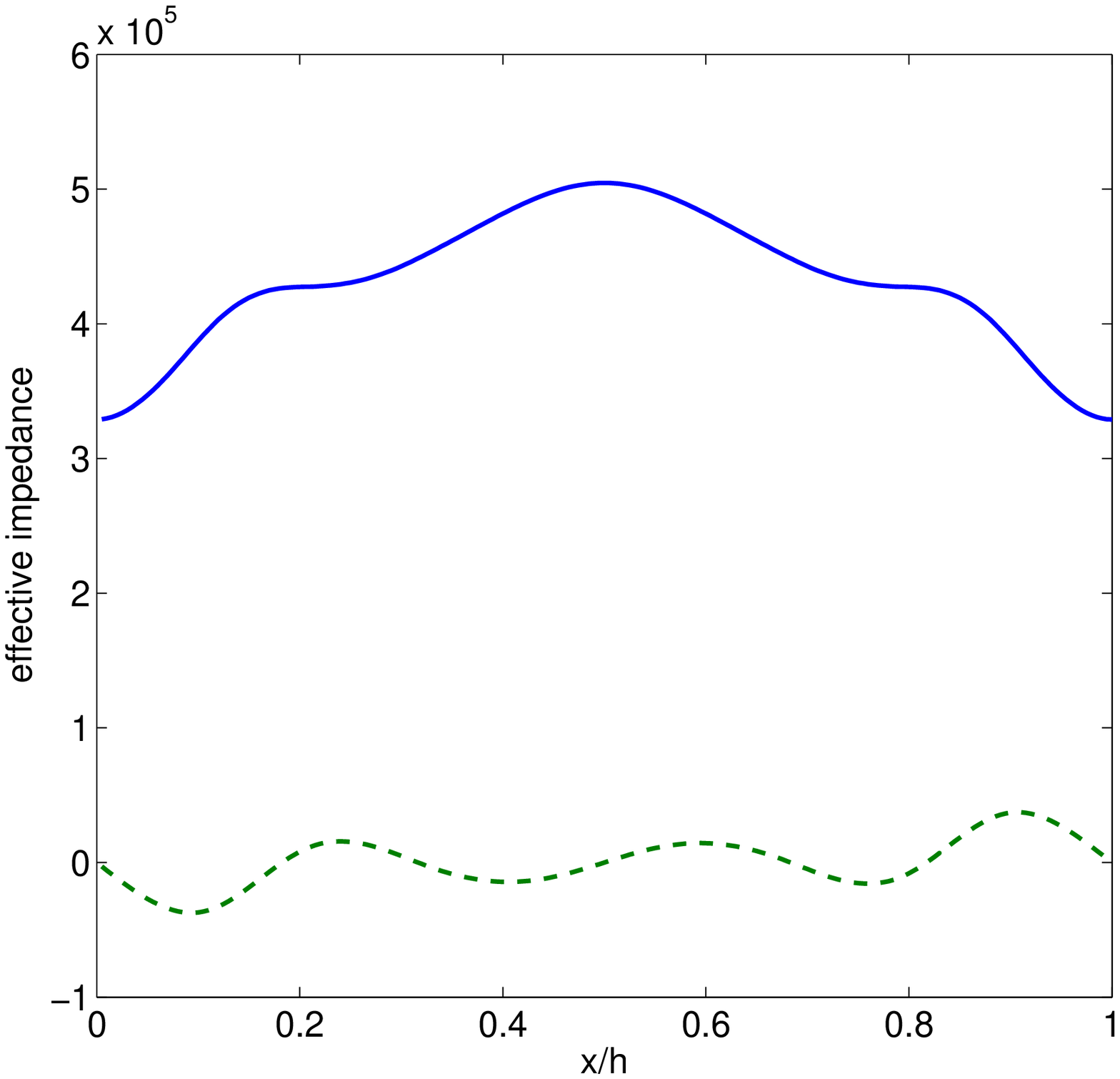}\\
(a)\hskip 6cm (b)
\caption{\small Effective impedance $Z^{eff}(X)$ (a) at frequency $f=17.5$\,kHz, (b) at $f=25$\,kHz.}
\end{figure}

\section{Discussion}
\setcounter{equation}{0}

The broad conclusion is that an ``effective medium'' description of a composite medium provides a
reasonable approximation for its response, so long as the predicted ``effective wavelength''
is larger than $2h$ -- say at least $2.5h$. This is consistent with the findings of Srivastava
and Nemat-Nasser \cite{SNN13} though they expressed it somewhat differently. Here, $h$ represents the
period of the microstructure. For a more general random medium, a similar limitation is to be expected;
an order of magnitude can be given for $h$ but it is not defined precisely. Of course, the conclusion,
even for a periodic medium, is a tentative one, being based on only two simple examples.

A periodic medium throws up a very specific difficulty as soon as the effective wavelength is
double the period. The imaginary part of the parameter called $\mu h$ in this work is mathematically
determined only up to a multiple of $2\pi$. While it is entirely natural that $\mu h$ should tend to
zero as frequency tends to zero, the choice of phase into and beyond the first stop band becomes
problematic. Perhaps the most ``natural'' choice is that $\mu h$ should be a continuous function of
frequency. The calculated phase speed then remains positive, even though the visually perceived
phase speed may be negative, due to aliasing. The problem is a real one for effective property
calculation because the choice of phase affects the definition of the periodic parts of the fields
and their associated averaged values. The ``alternative view'' presented in Section 5 circumvents
this difficulty: the formulation of that section reproduces stop bands exactly, and remains useful
through and beyond the first stop band. The ambiguity in definition of the
dispersion relation appears precisely at the first stop band. Note from Fig. 11 that 
${\rm Im}(\mu h) = \pi$ in the first stop band, and was chosen to be greater than $\pi$
thereafter.

When time permits, I hope to do further investigations, for a material with a stronger resonance
at a low frequency for which ``effective medium'' theory should be more accurate, and try to throw
some additional light on phenomena associated with apparently negative effective modulus and density.

\end{document}